\documentclass[published]{JHEP3} 
\JHEP{08(2003)056}
\received{June 10, 2003}
\accepted{August 29, 2003}

\skip\footins = 1\bigskipamount plus 2pt minus 4pt

\usepackage{epsfig}

\newcommand{\sfrac}[2]{\frac{#1}{#2}}
\newcommand{\be}{\begin{equation}}
\newcommand{\ee}{\end{equation}}
\newcommand{\nn}{\nonumber}
\def\bea{\begin{eqnarray}}
\def\eea{\end{eqnarray}}
\def\ansatz{\emph{ansatz}}
\def\ansatze{\emph{ans\"atze}}
\def\exd{{\rm d}}
\def\pref#1{(\ref{#1})}
\def\M{\mathcal M}

\def\g{{\rm g}}
\newcommand\SO{\mathop{\rm SO}\nolimits} 
\newcommand\CSO{\mathop{\rm CSO}\nolimits} 
\newcommand\ISO{\mathop{\rm ISO}\nolimits} 
\newcommand\U{\mathop{\rm {}U}\nolimits} 
\newcommand\SL{\mathop{\rm SL}\nolimits} 
\newcommand\SU{\mathop{\rm SU}\nolimits} 

\title{General brane geometries from scalar potentials: gauged
  supergravities and accelerating universes}

\author{Clifford P. Burgess\\
Physics Department, McGill University,  3600 University Street\\
Montr\'eal, Qu\'ebec, Canada, H3A 2T8\\
E-mail: \email{cliff@physics.mcgill.ca}}

\author{Carlos N\'u\~nez\\
Center for Theoretical Physics, Massachusets Institute of Technology\\
Cambridge, MA 02139, USA\\
E-mail: \email{nunez@lns.mit.edu}}

\author{Fernando Quevedo and Ivonne Zavala C.\\
Centre for Mathematical Sciences, DAMTP, University of Cambridge\\
Cambridge CB3 0WA UK\\
E-mail: \email{f.quevedo@damtp.cam.ac.uk}, \email{E.Zavala-Carrasco@damtp.cam.ac.uk}}

\author{Gianmassimo Tasinato\\
Physikalisches Institut der Universit\"at Bonn, \\
Nussallee 12, 53115 Bonn, Germany\\
E-mail: \email{tasinato@th.physik.uni-bonn.de}}

\abstract{We find broad classes of solutions to the field equations
for $d$-dimensional gravity coupled to an antisymmetric tensor of
arbitrary rank and a scalar field with non-vanishing potential.  For
an exponential potential we find solutions corresponding to brane
geometries, generalizing the black $p$-branes and S-branes known for
the case of vanishing potential. These geometries are singular at the
origin with up to two horizons.  When the singularity has negative
tension or the cosmological constant is positive we find
time-dependent configurations describing accelerating
universes. Special cases give explicit brane geometries for gauged
supergravities in various dimensions, and we discuss their
interrelation. Some examples lift to give new solutions to 10D
supergravity. Limiting cases preserve a fraction of the
supersymmetries of the vacuum. We also consider more general
potentials, including sums of exponentials. Exact solutions are found
for these with up to three horizons, with potentially interesting
cosmological interpretation. Further examples are provided.}

\begin{document}

\section{Introduction}\label{part1}

There has been considerable study of the solutions to Einstein's
equations in the presence of the matter fields which arise within
\pagebreak[3] supergravity theories in four and higher
dimensions. Such solutions have been invaluable to the development of
our understanding of string theory, such as through the identification
of the extremal limit of static black $q$-brane solutions of type-II
supergravities with D-branes~\cite{pol}.

Although initial interest was restricted to static and/or
supersymmetric solutions, late attention has turned towards
constructing time-dependent solutions for the
scalar-tensor-gravity system, which have been suggested to describe
transitions between different vacua within string
theory~\cite{gutperle}. In particular, the decay process of unstable
D-branes or D-brane--anti-D-brane pairs, such as described by the
dynamics of open string tachyons, may be related at long wavelengths
to time-dependent supergravity solutions called
S-branes~\cite{lukas,sbranes1,sbranes2}, which are extended,
space-like, solitonic objects embedded within time-dependent
backgrounds.

There are equally interesting potential applications to cosmology.
For instance solutions exist describing time-dependent universes
having contracting and expanding phases which are separated by
horizons from static regions containing time-like singularities.
Since observers in these universes can travel from the contracting to
expanding regions without encountering the usual space-like
singularities, they might ultimately lead to bouncing cosmologies.
Alternatively, their expanding regions may describe accelerating
cosmologies, a subject that recently has received much attention from
the string community~\cite{townsend2003,accelerating}~\footnote{Notice
  that the solutions presented in~\cite{townsend2003} are particular
  cases of solutions given in~\cite{sbranes1,ohta}.}.  These solutions
appear to evade the usual singularity or no-go theorems because of the
negative tension of the time-like singularities which
appear~\cite{gqtz}--\cite{cc}.

The exact solutions which have been constructed to date
use the field content motivated by the bosonic spectrum of low-energy
string theory in ten dimensions --- consisting of the metric tensor
$g_{\mu\nu}$, a dilaton scalar, $\phi$, and the field strength,
$F_{q+2}$, of an antisymmetric gauge form having rank $q+1$, which
might arise from either the NSNS or the RR sector, and which is
conformally coupled to the dilaton. For all these solutions the
dilaton field is assumed to have no potential, such as is typically
found for the simplest low-energy string configurations.

Our goal in the present paper is to extend these analyses to
supergravity systems having nontrivial scalar potentials. We are
motivated to do so because any real application of these solutions to
low-energy applications (like cosmology) is likely to require a
nontrivial potential for the dilaton, as well as for the other
low-energy moduli. These potentials are expected to be generated in
the full theory by a combination of non-perturbative effects and
compactification. Indeed, several well-defined scalar potentials arise
in gauged supergravities that are derived as compactified string
theories with non-vanishing background fluxes.  An alternative
motivation is also the study of the evolution of tachyon fields, such
as the open-string tachyon arising in brane-antibrane annihilation,
which for a single tachyon can be written as a scalar having a
nontrivial potential.

In particular, we present a systematic procedure for constructing
solutions of the field equations including non-vanishing
potentials. The procedure relies on introducing an \ansatz\ which
allows us to reduce the problem of solving the full field equations to
the integration of a single nonlinear ordinary differential equation,
whose form depends on the form of the scalar potential. This equation
has solutions which permit simple analytic expressions for potentials
of the generalized Liouville type 
\be
\label{eqintro} 
V\, =\,
\sum_{i=1}^L \Lambda_i e^{-\lambda_i \phi}\,, 
\ee 
such as often arise in explicit supergravity compactifications. We
therefore focus most of our attention to potentials of this type, and
in particular to the simplest case where the sum is limited to only
one term. (We return, however, to some more general examples in the
last subsections of the paper.)

Since the paper is quite long, most readers won't wish to read it
straight through from front to back. We therefore provide here a broad
road map of its layout, to facilitate better browsing. To this end we
divide the remainder of our discussion into four parts. Counting the
introduction you are now reading as section~\ref{part1}, the others
are:

\paragraph{Section~\protect\ref{part2}:}
here we describe the simplest example which illustrates our mechanism
of generating solutions to the Einstein/$(q+1)$-form potential/dilaton
system in $q+ n + 2$ dimensional spacetime. We do so by the choosing
an \ansatz\ for the metric, for general $V(\phi)$, from which
considerable information may be drawn about the resulting geometry. We
then specialize to the Liouville potential given by~(\ref{eqintro})
with a single term in the sum. This provides the simplest context
within which to see how our construction works, and it already
contains many supergravities of practical interest. For this potential
we present broad new classes of geometries describing brane-like
configurations, which for special choices of parameters reduce to
those solutions which are already in the
literature~\cite{wilt1}--\cite{zhang}.

For these solutions we study the global properties of the geometry,
and describe non-standard static brane backgrounds as well as new
examples of time-dependent backgrounds having a rich global space-time
structure. Many of their features follow from the study of their
asymptotic behavior which, due to the non-standard form of the
solution we find for the scalar field, depends both on the
cosmological constant\footnote{For convenience we call the constant
  $\Lambda$ in the Liouville potential the `cosmological constant',
  even though it describes a term in the potential which is not a
  constant in the Einstein frame.} and on the conformal couplings of
the dilaton to the various other fields. By studying various limits we
find smooth connections with well-known asymptotically-flat
geometries, including both static (Schwarzschild,
Reissner-Nordstr\"om) and time-dependent (S-brane--like)
configurations.

Among our static solutions we find geometries having the same global
structure as have Schwarzschild, Reissner-Nordstr\"om, and AdS
space-times. The time-dependent solutions include those which share
the same global properties of the S-brane solutions described
in~\cite{gqtz,gutperle}, as well as of de Sitter-like space-times (but
with a time-like singularity at the origin), and of de
Sitter-Schwarzschild space-times. Our solutions contain examples of
time-dependent, asymptotically-flat geometries for any choice of
$\Lambda$, and for a range of the conformal coupling, $\lambda$. The
time-dependence of these solutions can be interpreted as being due to
the presence of negative-tension time-like singularities, generalizing
in this way the geometrical interpretation of~\cite{bqrtz}.

We emphasize that the cosmological solutions we obtain are equally
valid for either sign of the cosmological constant. It is also
noteworthy that, similar to the geometries studied in~\cite{gqtz}, we
find solutions which describe accelerating
universes~\cite{gqtz,bqrtz,paul}, for which the Ricci curvature
vanishes at infinity due to the presence of the scalar field. This is
reminiscent of some quintessence models, for which a slowly-rolling
scalar field accelerates the expansion of the universe in a way which
decreases to zero at asymptotically late times (in contrast with a
pure cosmological constant). In this sense we furnish exact
supergravity solutions which share key features of quintessence
cosmologies, and which can provide a good starting point for more
detailed cosmological model building.

\paragraph{Section~\protect\ref{part3}:}
in this subsection we specialize the general discussion of the
Liouville potential in section~\ref{part2} to the specific parameters
with which this potential arises in gauged and massive supergravities
in various dimensions. We do so in order of decreasing dimension,
starting with the massive type-IIA supergravity in ten dimensions,
including a positive cosmological constant such as has been
interpreted in~\cite{pol} as being due to the $vev$ of the $F_{10}$
field strength which is naturally present in the RR sector of type-IIA
superstrings.

We continue to consider lower-dimensional gauged supergravities, for
which the presence of a cosmological constant is a byproduct of the
compactification procedure, being generated by dimensional reduction
on compact (spherical or toroidal) or non-compact (hyperbolic)
spaces. Our general methods applied to these specific examples give a
variety of exact solutions. Some of these are already known in the
literature, mainly in the form of domain walls which preserve section
of the supersymmetries of the vacuum.  The importance of these
configurations for the study of suitable generalizations of the
AdS-CFT correspondence has been noticed in~\cite{boonstra}, and is
currently a field of intense study.

Our methods produce charged black brane (hole) solutions for many
examples of gauged supergravities and, in general, supersymmetric
domain-wall solutions are obtained from these by appropriately setting
some of the parameters to zero. Interestingly, we also find
cosmological configurations for some of the solutions by choosing the
parameters differently. Moreover, we use general oxidation methods to
lift the lower-dimensional geometries to exact solutions of a
ten-dimensional theory which describes black branes wrapped about
various manifolds in ten dimensions. This lifting procedure to 10
dimensions in some cases can provide connections amongst the various
lower-dimensional solutions.  Related black hole solutions within
gauged supergravities have been studied in~\cite{behrndt,cvetic99}.

\paragraph{Section~\protect\ref{part4}:}
this section of the paper presents the procedure which allows the
generation of solutions for more general scalar potentials. Here we
derive our results starting from metrics depending on two coordinates,
following a treatment in five dimensions by Bowcock et
  al.~\cite{bowcock} and subsequent workers~\cite{langlois}. This
method allows us to identify the explicit functional dependence of the
metric and dilaton for general potentials. Using this technique we
reduce the problem of finding a brane solution to that of solving a
single nonlinear ordinary differential equation for the scalar field.

This differential equation may be solved numerically given any
potential, but may be solved analytically for specific kinds of
potentials. We provide examples of this by deriving field
configurations which solve the field equations when the scalar
potential is the sum of up to three exponential terms. We also
provide a few more complicated examples, such as for a flat,
vanishing-charge geometry, with potential given by:
\be
    V(\phi)\ =\ e^{b \phi^2} (a-b\phi^2) \, .
\ee

Finally, we end in section~\ref{part5} with some general comments
about our results and on possible generalizations.

\section{The simplest case}\label{part2}

We begin by describing the equations we shall solve, as well as
presenting our solutions within their simplest context. Our main focus
in this subsection is an exponential potential, but we first proceed
as far as possible without specifying the potential explicitly. A more
systematic way of generating solutions from a general potential is
described in section~\ref{part4}, below.

\subsection{The set-up}\label{action}

\paragraph{The  action.}
Consider the following action in $(q+n+2)$ dimensions, containing the
metric, $g_{\mu \nu}$, a dilaton field, $\phi$, with a general scalar
potential, $V(\phi)$, and a $(q+2)$-form field strength, $F_{q+2} =
\exd A_{q+1}$, conformally coupled to the dilaton:
\be
    \label{generalaction}
S= \int_{{\cal M}_{q+n+2}} d^{q+n+2}x \sqrt{|g|} \left[ \alpha
    {\mathcal R} - \beta
(\partial \phi)^2 - \frac{\eta}{(q+2)!} {\rm e}^{-\sigma \phi} F^2_{q+2}
             -V(\phi) \right] .
\ee
Here ${\mathcal R}$ is the Ricci scalar built from the metric, and we
use MTW conventions as well as units for which $8 \pi G_{d}=1$, where
$G_{d}$ is the higher dimensional Newton constant.

Stability requires the constants $\alpha, \beta$, and $\eta$ to be
positive and, if so, they may be removed by absorbing them into
redefinitions of the fields. It is however useful to keep $\alpha$,
$\beta$ and $\eta$ arbitrary since this allows us to examine the cases
where each constant is taken to zero (to decouple the relevant
fields). For specific values of the parameters this action can be seen
as section of the low-energy string theory action, including a
potential for the dilaton.

In this subsection our main application is to the Liouville potential,
$V = \Lambda \, e^{-\lambda \phi}$, for which Wiltshire and
collaborators~\cite{wilt2} have shown that the equations of motion do
not admit black hole solutions except for the case of a pure negative
cosmological constant, $\lambda = 0$ and $\Lambda < 0$. Their
arguments assume the fields do not blow up at infinity, and this is
the condition we relax in order to find solutions. In particular we
entertain scalar fields which are not asymptotically constant, but
which can diverge at infinity at most logarithmically.

\paragraph{The equations of motion.} 
The field equations obtained for the action of
eq.~(\ref{generalaction}) are given by:
\be  
\left\{\begin{array}{l}
\displaystyle  \alpha G_{\mu\nu}
= \beta T_{\mu \nu}[\phi] + \frac{\eta}{(q+2)!} {\rm e}^{- \sigma \phi} T_{\mu
\nu}[F_{q+2}] - \frac{1}{2}V(\phi) g_{\mu\nu} 
\\[8pt] 
\displaystyle 
 2\,\beta\,\nabla^2 \phi =-\sigma \frac{\eta}{(q+2)!} \, {\rm e}^{-\sigma\phi}
 F^2_{q+2}  + \frac{d}{d \phi} V(\phi)   \,, 
\\ [8pt] 
\displaystyle 
 \nabla_{\mu} \left( {\rm e}^{-\sigma\phi} F^{\mu \cdots}\right) = 0\,, 
\end{array} 
\right.
\ee
where
$$
T_{\mu \nu}[\phi] = \nabla_\mu \phi \nabla_\nu \phi
-\sfrac{1}{2} g_{\mu \nu} (\nabla\phi)^2 \quad {\rm and} \quad
T_{\mu \nu}[F_{q+2}] = (q+2) F_\mu{}^{\cdots}
        F_{\nu\cdots} - \sfrac{1}{2}g_{\mu\nu} F^2_{q+2}\,.
$$

Since our interest is in the fields generated by extended objects
charged under the $(q+1)$-form potential, we look for solutions
having the symmetries of the well-known black $q$-branes. To this
end we consider the following metric \ansatz:
\be
    ds^2 =  -h(\tilde r) dt^2 +  h(\tilde r)^{-1} d{\tilde r}^2
    + \tilde f^2(\tilde r) dx_{k,n}^2
             + \tilde g^{2}(\tilde r) dy_{q}^2\,,
\ee
where $dx_{k,n}^2$ describes the metric of an $n$-dimensional
maximally-symmetric space with constant curvature $k=-1,0,1$ and
$dy_{q}^2$ describes the flat spatial $q$-brane directions. In
addition we take the $q$-brane charge to generate a $(q+2)$-form field
which depends only on $\tilde r$ and which is proportional to the
volume form in the $(t,\tilde r,y)$ directions, and we assume a
$\tilde r$-dependent dilaton field.

With these choices the field equations for the antisymmetric
tensor imply
\be F^{{\tilde r}ty_1\cdots y_q} = \frac{Q
e^{\sigma\phi}}{\tilde f^n \tilde g^q}\, \epsilon^{{\tilde
r}ty_1\cdots y_q} \, , 
\ee
and the dilaton and Einstein equations reduce to the following
system
\be 
2\beta \tilde f^{2n}\, \left[\tilde f^n\tilde
g^qh\phi'\right]'=\tilde f^n \tilde g^q\left(\sigma\eta Q^2
e^{\sigma\phi} + \tilde f^{2n}\frac{dV}{d\phi}\right) , 
\ee
and
\bea 
n\, \frac{\tilde f''}{\tilde f}+ q\, \frac{\tilde
g''}{\tilde g} & = &
-\frac{\beta}{\alpha}\left( \phi' \right)^2 
\nn \\
-\alpha \tilde f^n\, \left[h \tilde g^q \left(\tilde
f^n\right)'\right]' & = & -n(n-1)\alpha k {\tilde f}^{2n-2} \tilde
g^q  + \frac{n}{d-2} V(\phi) \tilde f^{2n} \tilde g^q
+ \frac{\eta \, n (q+1)}{d-2}  Q^2 \tilde g^q
e^{\sigma\phi}
\nn \\
\alpha (d-2) \tilde f^n \left[ h \tilde f^n \left(\tilde
g^q\right)'\right]' & = & \eta q (d-3-q)  Q^2 g^q e^{\sigma
\phi}- q \tilde f^{2n} \tilde g^q \, V(\phi) \,.
\eea
We arrive at a system of four differential equations for the four
functions $h,\tilde f,\tilde g,\phi$, whose solution requires a
specification of the explicit form for the potential $V(\phi)$.

It is possible, however, to go a long way without needing to specify
the potential if we make a simplifying \ansatz\ for the components of
the metric. To this end let us assume the metric component $g$ can be
written in the form
\be
 \tilde g= r^c
\ee
for constant $c$, and with the new variable $r$ defined by the
redefinition
\be 
r=  \tilde f(\tilde r) \, .
\ee
It is also convenient to think of the dilaton as being a logarithmic
function of $r$, with
\be 
\label{dilaton} 
\phi(r) = \M \, S(\ln r) \,, 
\ee
where $\M$ is a constant (whose value is given explicitly below).

\paragraph{The solutions.}
Subject to these \ansatze\ the solutions to the previous system of
equations are given by
\bea\label{genformet}
ds^2 &=&  -h(r) dt^2 +  \frac{dr^2}{g(r)} + r^2 dx_{k,n}^2
+ r^{2c} dy_{q}^2\,, 
\\
F^{try_1\dots y_q} &=& {Q}\, e^{\sigma \M S(\ln r)-L(\ln r) } r^{  -\M^2-(N-1)}
 \, \epsilon^{try_1\dots y_q} \,,
\label{qform}
\eea
with
\be
\label{ge} 
g(r)= h(r) \, r^{-2(N- 1)}\, e^{-2L(\ln r)}\,, 
\ee
and the function $L(\ln r)$ is given in terms of $S(\ln r)$ by
\be 
{d L \over dx}(x) = \frac{\beta}{\alpha}\ \left( {dS \over dx}
(x) \right)^2 .
\ee
Finally, the constants $\M$ and $N$ are related to the parameters
$n$, $q$ and $c$ by
\be
\M^2=n+cq \,,\qquad N=\frac{n+c^{2}q}{\M^{2}}\,.
\ee

So far we have not had to give the form for the scalar potential, but
this cannot be avoided if the remaining unspecified functions $S(\ln
r)$ and $h(r)$ are to be obtained explicitly. In principle, once
$V(\phi)$ is given these remaining functions may be found by solving
the remaining field equations. Before doing so it is instructive first
to extract as much information as we can about the geometries which
result in a potential-independent way.

Our \ansatze\ allow the following general conclusions to be drawn:
\begin{enumerate}
\item{}The solutions describe a flat $q$-dimensional extended object
  or $q$-brane. In particular the metric of eq.~(\ref{genformet}) has
  the symmetry $\SO(1,1) \times O_{k}(n) \times \ISO(q)$, where
  $O_{k}(n)$ refers to $\SO(1,n-1)$, $\ISO(n)$ or $\SO(n)$ for
  $k=-1,0,1$ respectively. \footnote{We shall see that in specific
    static cases this symmetry can be enhanced for special choices for
    some of the parameters. In particular, we find examples for which
    the isometry group is promoted to $\SO(1,n) \times \ISO(q+1)$.}
\item{}The coordinates used break down for those $r$ which satisfy
  $h(r)=0$. These surfaces correspond to regular horizons rather than
  to curvature singularities. Their number is given by the number of
  sign changes in $h(r)$, which is one or two for most of the cases we
  discuss below.
\item{}Typically the limit $r \to 0$ represents a real singularity,
  and often describes the position of the extended object which
  sources the geometry. The charge and tension of this source may be
  read off from the fields it generates, just as for Gauss' Law in
  electromagnetism. For the $(q+1)$-form gauge potential this leads to
  charge $Q$.
\item{}The tension of the source may be determined using the Komar
  formalism, following the methods discussed in~\cite{bqrtz}. In
  general the result depends on the radius at which the fields are
  evaluated, since the gravitational and other fields can themselves
  carry energy. The \emph{tension} calculated using the
  metric~(\ref{genformet}) at the space-like hypersurface at fixed
  $r$, turns out to be given by
\be
\label{tension} 
\mathcal T = \frac{C \alpha
\,V_{n+q}}{2}\,r^{n+cq} \sqrt{\frac{g}{h}} \,h'(r)\,, 
\ee
where $C$ is a normalization constant and $V_{n+q}$ is the volume of
the $(n+q)$-dimensional constant-$r$ hypersurface over which the
integration is performed.

When the geometry is static for large $r$, the tension calculated in
this way tends to the ADM mass as $r \to \infty$ provided one chooses
$C=4$. For solutions for which the large-$r$ regime is time-dependent
there are typically horizons behind which the geometry is static, and
this formula can be used there to calculate the tension as a function
of position.

\item{}Given the existence of horizons it is possible to formally
  associate a ``temperature'' with the static regions by identifying
  the periodicity of the euclidean subsection which is nonsingular at
  the horizon (see ref.~\cite{bqrtz}). For the metric of
  eq.~(\ref{genformet}) the result is \be\label{temperature} T =
  \sqrt{\frac{g(r_{h})}{h(r_{h})}} \frac{h'(r_h)}{4\pi}\,, \ee where
  $r_{h}$ corresponds to the horizon's position. Notice that
  eq.~\pref{ge} ensures that $g(r_h)/h(r_h)$ is nonsingular even
  though $h(r_h)$ vanishes. An entropy associated with this
  temperature can also be computed in the same fashion as was done
  in~\cite{bqrtz} with a similar result.
\item{}The Ricci scalar is given by the comparatively simple
  expression
\be
\label{Ricci}
\mathcal R = g(r)\frac{(\mathcal M\,S'(\ln r))^2}{2\,r^2} + \frac{(n+q+2)
V(\phi)}{n+q} + \frac{(q+2-n)\,\eta \,Q^2 h(r)}{(n+q)\, g(r)}
e^{\sigma\mathcal M S(\ln r)}r^{2(1-n-N)} \,.
\ee
With this expression it is straightforward to check how the Ricci
scalar behaves at infinity. In particular, we shall find that
$\mathcal R$ vanishes asymptotically for many solutions,\footnote{It
  does so for solutions in classes I, II, and III below.} when we have
a non-constant scalar field with a Liouville potential.

\item{}An arbitrary constant can always be added to the functional
  form of $\phi$, as this can always be absorbed into the other
  constants. When the potential has more than one term and it has an
  extremum, there will be an automatic solution corresponding to
  standard dS or AdS, depending on the sign of the potential at the
  critical point.
\end{enumerate}

It is important to stress that all the above conclusions may be drawn
\emph{independently} of the choice of the potential.

To proceed further, however, we must choose a particular form for
$V(\phi)$. In the next few subsections we specialize to the Liouville
potential
\be
\label{lform} V(\phi)\ =\ \Lambda e^{-\lambda\phi}\,, 
\ee
which has the twin virtues of being simple enough to allow explicit
solutions and of being of practical interest due to its frequent
appearance in real supergravity lagrangians. We return in
section~\ref{part4} to more general choices for $V(\phi)$. As we shall
see in the next subsection this simple potential is already rich
enough to provide geometries having many interesting global
properties.

For the Liouville potential the Einstein equations determine the
dilaton function, $S(\ln r)$, to be:
\be
\label{Ssinliu} 
S(\ln r)= \rho \ln r\,. 
\ee
where the parameter $\rho$ is a proportionality constant to be
determined. This form in turn implies $L(\ln r)\ =
\frac{\beta}{\alpha} \rho^2 \ln r$. Using these expressions in the
remaining field equations in general implies the function $h(r)$ is
over-determined inasmuch as it must satisfy two independent
equations. The next subsection shows in detail how these equations
admit solutions of the form:
\be 
h(r) = \sum_{j=1}^4 \alpha_j r^{a_j} 
\ee
where the coefficients $\alpha_i$ and $a_i$ are determined in terms of
the parameters $k, Q^2, \Lambda_i$ and an integration constant, $M$,
to be defined below.

\subsection{Explicit brane solutions for single Liouville potential}
\label{123456}

In this subsection, we present four classes of solutions for the
Liouville potential~(\ref{lform}), with $\Lambda\neq 0$.

Let us start by rewriting the general form of the solutions in
this case, substituting in~(\ref{genformet}) and~(\ref{ge}) the
form of $S$ given by formula~(\ref{Ssinliu}). We find in this way:
\bea
ds^2 &=&  -h(r) dt^2 +  \frac{dr^2}{g(r)} + r^2 dx_{k,n}^2
             + r^{2c} dy_{q}^2\,, 
\\
\label{dilatonL}
\phi(r) &=&\rho \, \M \, \ln r\,, 
\\
\label{qformL}
F^{try_1\dots y_q} &=& Q \,r^{ \sigma\rho \M
       -N-\M^{2} -\beta\rho^2/\alpha+1} \, \epsilon^{try_1\dots y_q} \,,
\eea
with
\be\label{geL} g(r)= h(r) \, r^{-2(N+ \beta\rho^2/\alpha- 1)}\, .
\ee

With these expressions the $(tt)$ and $(rr)$ components of
Einstein's equations imply the following condition for $h$:
\bea \label{solh2}
\M^2 \,h(r) &=& \left[ \frac{n(n-1)\,k}
{(\M^2 - 2 + N + \beta\rho^2/\alpha)}\right]
r^{2(\beta\rho^2/\alpha+N-1)} -
2 M \,\M^2 r^{N+\beta\rho^2/\alpha-\M^{2}  } -
\nonumber  \\
&& - \left[ \frac{\eta Q^2}{\alpha
(\sigma\rho \M - 2n +\M^{2}+N+ \beta\rho^2/\alpha)} \right]
 {r^{\sigma\rho \M + 2\beta\rho^2/\alpha}\over r^{2(n-N)}} -
\nonumber  \\
&& - \left[ \frac{ \Lambda}
{\alpha(\M^2 + N + \beta\rho^2/\alpha - \lambda \rho \M])}\right]
\frac{r^{2(N+\beta \rho^{2}/\alpha)}}{r^{\lambda \rho \M}}\, ,
\eea
where $M$ is an integration constant. On the other hand the dilaton
equation implies $h(r)$ must also satisfy:
\bea
\label{solh1}
\M\,h(r) &=&  - 2 M \,\M \, r^{N+\beta\rho^2/\alpha-\M^{2}}
{+} \left[ \frac{\eta \,\sigma\,Q^2}{2\beta\rho
(\sigma\rho \M  -2n +  \M^2+N+ \beta\rho^2/\alpha)} \right]
 {r^{\sigma\rho \M + 2\beta\rho^2/\alpha}\over r^{2(n-N)}}{-}
\nonumber  \\
&&- \left[ \frac{\lambda \Lambda}
   {2\beta\rho(\M^2 + N+ \beta\rho^2/\alpha -
  \lambda \rho \M)}\right]
  \frac{r^{2(N+ \beta \rho^{2}/\alpha)}}
           {r^{\lambda \rho \M}}\,.
\eea
The $(y_q y_r)$ components of the Einstein equations impose the
further conditions
\bea
\label{constrayy}
\lefteqn{q\cdot (c-1)\left[ \frac{n(n-1)\,k}{\M^2 } 
- \frac{\eta Q^2}{\alpha \M^{2}}
r^{\sigma \rho \M-2(n-1)} - \frac{\Lambda}{\alpha \M^{2}}
r^{-\lambda \rho \M+2} \right]=}\hspace{15em} && 
\nn \\
&=& q \cdot \left[(n-1)\,k- \frac{\eta Q^2}{\alpha}
 r^{\sigma \rho \M-2(n-1)}\right] .\qquad
\eea
We write explicit factors of $q$ on both sides of this last equation
to emphasize that these equations only exist when $q \ne 0$.

In order to obtain solutions we must require that eqs.~(\ref{solh2})
and~(\ref{solh1}) imply consistent conditions for $h(r)$, and we must
also impose eq.~(\ref{constrayy}). We find these conditions can be
satisfied by making appropriate choices for the parameters in the
solutions. We identify four classes of possibilities which we now
enumerate, giving interesting solutions for extended objects.

\begin{description}
\item[Class I.] Here we choose $k=0$ and match the coefficients of
  $\Lambda$ and $Q^{2}$ in formulas~(\ref{solh2}) and~(\ref{solh1}).
  For $q=0$, this class corresponds to the one considered
  in~\cite{cai}. For $q \neq 0$, the condition~(\ref{constrayy})
  implies that $c=1$ and $Q=0$ in order to have brane-like solutions.
\item[Class II.] In this case we take $k \neq 0$ and identify the
  exponents of $r$ in the terms proportional to $\Lambda$ and $k$
  in~(\ref{solh2}), allowing these two terms to be merged
  together. Next, we identify the two remaining terms in~(\ref{solh2})
  with the two terms of~(\ref{solh1}). For $q=0$ we need not
  impose~(\ref{constrayy}) and the resulting solutions can have
  nonzero $Q$~\cite{mann,cai}. By contrast, for $q \ne 0$
  eq.~(\ref{constrayy}) implies $Q=0$ (but $c \ne 1$).
\item[Class III.] Here we demand that the terms proportional to $Q^{2}$
  and $k$ in~(\ref{solh2}) share the same power of $r$ and so combine
  together. If $q \ne 0$ then we also impose eq.~(\ref{constrayy}),
  which implies $c=1$. For $q=0$ the solutions have the same form, but
  without the constraint coming from~(\ref{constrayy}).
\item[Class IV.] In this case we ask to fuse together the terms
  proportional to $k$, $\Lambda$, and $Q^{2}$ in~(\ref{solh2}) (and
  the same for the terms proportional to $\Lambda$ and $Q^{2}$
  in~(\ref{solh1})), by asking them to share the same power of $r$. We
  then identify the remaining term of~(\ref{solh2}) with the one
  of~(\ref{solh1}).  When $q \neq 0$, we impose
  also~(\ref{constrayy}).
\end{description}

Naturally, solutions belonging to different classes above can coincide
for some choices of parameters. We now present these solutions in more
detail. For brevity we display explicitly only the form of the metric
coefficient $h(r)$ for the solutions, since the expressions for the
function $g(r)$, the scalar $\phi$, and the antisymmetric forms are
easily obtained using
formulae~(\ref{dilatonL}),~(\ref{qformL}),~(\ref{geL}).

\paragraph{Class I.}
This class contains solutions only for $k=0$ (that is, for flat
maximally-symmetric $n$-dimensional submanifolds).
To obtain solutions we must also impose the following relations
among the parameters:
\be
\left\{
\begin{array}{l}
\alpha \sigma \M = -2 \beta \rho \,, \\ [3pt]
\alpha \lambda \M = 2 \beta \rho  \,, \\ [3pt]
c= 1 \quad {\rm and}  \quad  Q =0 \quad {\rm if} \quad q \ne 0\,.
\end{array} \right.
\ee
The first two of these imply the solution only exists of the
lagrangian couplings satisfy $\sigma = - \lambda$. They also determine
the parameter $\rho$ of the dilaton \ansatz\ in terms of these
couplings: $\rho = \alpha \lambda \M/(2 \beta)$. Moreover, if $q\neq
0$, we must also require $c=1$ and $Q=0$ as specified by the last
condition. For $q=0$, $Q$ need not vanish. For either choice of $q$ we
have $N=1$.

With these choices the metric function $h$ becomes
\be
\label{class1}
h(r) = -2 M r^{1- \M^{2} +\beta\rho^2/\alpha} -
 \frac{\Lambda
 r^{2}}{\alpha \,\M^{2}\,[\M^{2}{-}\beta\rho^2/\alpha{+}1]} 
-\frac{\eta\,Q^2}{\alpha\,\M^{2}[\M^{2} {-}2n
                               {-}\beta\rho^2/\alpha{+}1]\,r^{2(n-1)}}\,.
\ee

\paragraph{Class II.}
This class contains geometries having any curvature $k = 0, \pm
1$. The parameters of the \ansatz\ must satisfy the constraints:
\be 
\label{classii}
\left\{
\begin{array}{l}
\lambda \rho \M = 2 \,, \\ [3pt]
\alpha\sigma\M= -2 \beta \rho \,, \\ [3pt]
\displaystyle
\Lambda= \frac{\alpha \sigma \,n(n-1)\,k}{\sigma + \lambda}\,,\\ [8pt]
\displaystyle
q\cdot  c= \frac{-2n\lambda + n\sigma}{ q (\sigma +\lambda)
                        -\lambda\,n)}\cdot q \,, \\ [8pt]
q\cdot Q = 0 \,.
\end{array} \right.
\ee

Consistency of the first two of these equations implies the condition
$\alpha \sigma \lambda \M^2 + 4 \beta = 0$, and $\rho^2 =
-{\alpha\sigma}/{\beta\lambda}$. The function $h$ then becomes
\be
\label{class2} 
h(r)= -2 M r^{N - \M^{2}
+\beta\rho^2/\alpha} - \frac{\lambda\Lambda\,
r^{2(\beta\rho^2/\alpha+N-1)}}
{2\beta\rho\, \M[\M^{2}{-}2{+}\beta\rho^2/\alpha{+}N]} 
-\frac{\eta\, Q^2 \, r^{2(N-n)}}{\alpha \M^2 [\M^2 {+}N {-} 2n {-}
\beta\rho^2/\alpha ] }\,.
\ee

\pagebreak[3] 

\paragraph{Class III.}
This class of solutions is new --- to our knowledge --- and turns out
to be among the most interesting in our later applications. It allows
geometries for any $k$ and for any $q$, with the constraints
\be\label{class3}
\left\{
\begin{array}{l}
\sigma \rho\M = 2(n-1)\,,  \\ [3pt]
\alpha \lambda\M = 2 \beta \rho  \,, 
\\ [3pt]\displaystyle 
Q^2 = \frac{\alpha\lambda\, n(n-1)k}{\eta(\lambda + \sigma)}\,,
\\ [8pt]
c=1 \,
\\ [3pt]
q \cdot  \sigma = q \cdot \lambda (n-1) \,.
\end{array} \right.
\ee
Consistency of the first two equations implies $\alpha \sigma \lambda
\M^2 = 4 (n-1) \beta$, which for $c=1$ implies the constraint $\alpha
\sigma \lambda (n + q) = 4 (n-1) \beta$. (If $q \ne 0$ and $n \ne 1$
then $\sigma$ can be eliminated from this last condition using the
final equation of \pref{class3}, leading to the restriction: \hbox{$\alpha
\lambda^2 (n+q) = 4\beta$.})

The \ansatz\ parameter $\rho$ then satisfies $\rho^2
=\frac{\alpha\lambda (n-1)}{\sigma\beta}$ and $N=1$. When $Q \ne 0$
there are solutions only when $k$ is nonzero and shares the sign of
$(n-1)\lambda/(\lambda + \sigma)$. For vanishing charge, $Q=0$, on the
other hand, there are only solutions for $k=0$ (if $n(n-1)\lambda$
does not vanish).

The metric function $h$ of eq.~(\ref{genformet}) then becomes
\be
\label{intsol}
h(r)=    -2 M r^{1 -\M^{2} +\beta\rho^2/\alpha} -
\frac{\Lambda\,r^{2}}{\alpha \M^{2}[\M^{2}-\beta\rho^2/\alpha+1]}
+\frac{\sigma\eta \,Q^2\,r^{2\beta\rho^2/\alpha}}
{2\beta\rho\,\M\,[\M^{2}-1+\beta\rho^2/\alpha]}\,.
\ee

\paragraph{Class IV.}
We subdivide this class into two cases, corresponding to $k=0$ and
$k\neq 0$.

\begin{description}
\item[$\bullet$]
The constraints in this case are
\be\label{class4a}
\left\{
\begin{array}{l}
\displaystyle
\rho \M = \frac{2n}{(\sigma+\lambda)}\,,
\\ [3pt]
\displaystyle
\eta Q^{2}\left(\frac{\beta}{\alpha}+\frac{n}{\rho^{2}}
-\frac{\lambda \M}{2 \rho} \right)
+\Lambda\left(\frac{\beta}{\alpha}-
\frac{\lambda \M}{2 \rho} \right)=0 \,,
\\ [3pt]
\displaystyle
c = \frac{2\beta\rho \, (1+\alpha \M\, [1+n]) + \sigma\alpha \M}
{2\beta\rho \,(1+\alpha \M\, [1-q]) + \sigma\alpha \M } \,.
\end{array} \right.
\ee
The function $h(r)$ is then given by
\be
h(r)= -2 M r^{N - \M^{2} +\beta\rho^2/\alpha} -
\frac{\left[ \eta\, Q^{2} +\Lambda \right] \,
r^{2\beta\rho^{2}/\alpha + 2N -\lambda\rho\M}} {\alpha \M^2
[\M^{2}+N+\beta\rho^{2}/\alpha - 2n + \sigma \rho\M]} \, . \ee

\item[$\bullet$]
In this case, the constraints are more restrictive:
\be
\label{class4b}
\left\{
\begin{array}{l}
\displaystyle
\lambda \rho \M =2\,,
\\ [8pt]\displaystyle
\sigma \rho \M = 2(n-1)\,,
\\ [8pt]\displaystyle
\eta Q^{2}\left(\frac{\beta}{\alpha}
+\frac{n-1}{\rho^{2}} \right)+\Lambda\left(\frac{\beta}{\alpha}-
\frac{1}{\rho^{2}}\right)= \frac{n(n-1) \beta}{2} k+ \,, 
\\[8pt]\displaystyle
 q\cdot \left[ \alpha\, (n-1)\,k \,[n\,(c-1)- \M^2]- \eta \,Q^2\,
[c-1 - \M^{2}]= \Lambda \,(c-1) \right]   \,.
\end{array} \right.
\ee
Consistency of the first two requires $\sigma = \lambda (n-1)$, and
the parameter $\rho$ is given by $\rho = 2/(\lambda \M)$.

The function $h$ reduces to
\be
h(r)=-2 M r^{N - \M^{2} +\beta\rho^2/\alpha}
+\frac{\left[ \alpha\, n(n-1)\,k-\eta \,Q^{2}
-\Lambda \right]\, r^{2\beta \rho^{2}/\alpha +2N-2}}
  {\alpha \M^{2}[\M^{2}-2+ \beta\rho^{2}/\alpha + N]}\,.
\ee
\end{description}

\subsection*{Properties of the solutions}\label{specprop}

In this subsection, we describe the global properties of the
geometries we have obtained. In particular, since $r$ is the
coordinate on which the metric depends, whether the metric is
time-dependent or not hinges on whether or not $r$ is a time-like or
space-like coordinate. This in turn depends on the overall sign of
$h(r)$, and section of our purpose is to identify how this sign
depends on position in the space, and on the parameters of the
solution.

We shall find that the presence of the Liouville potential allows much
more complicated structure for the charged dilatonic geometries than
is obtained without a potential. In particular we wish to follow how
the geometry is influenced by the sign of the cosmological constant
$\Lambda$, by the size of the conformal couplings of the dilaton,
$\sigma, \lambda$, and by the curvature parameter, $k$, of the
$n$-dimensional submanifold.

The solutions to our equations with $V(\phi) = 0$ are studied in
ref.~\cite{bqrtz}, who found that the geometries having
$n$-dimensional $k = 1$ submanifolds are the well known static
$q$-brane configurations, with at most two horizons. By contrast,
those configurations with $k = 0$ or $-1$ subspaces describe
time-dependent geometries separated from static regions by
horizons. These have been argued to correspond to special cases of
S-brane configurations.

Here we find that the addition of the Liouville potential modifies
this earlier analysis in several interesting ways. In general, we find
that if $\Lambda < 0$ the background is usually static, regardless of
the choice of the curvature $k$ of the subspace.  There can be
interesting exceptions to this statement, however, for sufficiently
large conformal couplings to the dilaton. These exceptional solutions
are quite appealing for cosmological applications, since they
correspond to time-dependent backgrounds even in the presence of a
Liouville potential with negative sign (that is the typical situation
when one considers compact gauged supergravities).

If $\Lambda > 0$, on the other hand, we have examples of new
geometries which are de Sitter-like, although they are not
asymptotically de Sitter, due to the presence of the dilaton.  Instead
the Ricci scalar asymptotically vanishes, although in most cases the
Riemann tensor itself does not. These asymptotic geometries are
intriguing because in them observers experience a particle
horizon. This asymptotic joint evolution of the dilaton and metric
resembles what happens in quintessence cosmologies, and in this sense
our solutions may lead to connections between supergravity and
quintessence models. Among these asymptotically time-dependent
backgrounds, we find examples in which an event horizon hides a
space-like singularities from external observers. (This is an
important difference relative to the existing S-brane geometries,
which have time-like singularities from which asymptotic observers can
receive signals).

For the remainder of the discussion we analyze each of the first three
classes in turn. In each case we start with an initial overview of the
main global features and then specialize to the limit $\Lambda \to 0$,
allowing us to see how the new geometries relate to those which were
previously known. In some cases we find in this way generalizations of
the earlier S-brane configurations.  In other cases we instead find
less well-known static geometries.  We end each case by focusing on a
four dimensional example in detail, since this exposes the geometries
in their simplest and clearest forms. In the following, we also limit
our discussion to the cases $q=0$ (i.e.\ to zero-branes) and $M \ge
0$, although the relaxation of these assumptions is straightforward to
perform.

\paragraph{Class I.}
As is clear from the expression for $h(r)$, the global properties of
the geometry are therefore largely controlled by the value of the
parameter $\rho$, which itself depends on the value of the conformal
coupling, $\lambda$, of the dilaton in the scalar
potential.\footnote{We give a more complete account of the global
  properties of our geometries in \textref{appendix}{appendix}.}  We
now consider the various possibilities. When $0 < \beta\rho^2 <
\alpha\,(n+1)$ the options depend on the sign of $\Lambda$. If
$\Lambda < 0$, $r$ is a spatial coordinate and the geometry is
static. By contrast, it is time dependent for positive $\Lambda$.

\FIGURE[t]{\epsfig{file=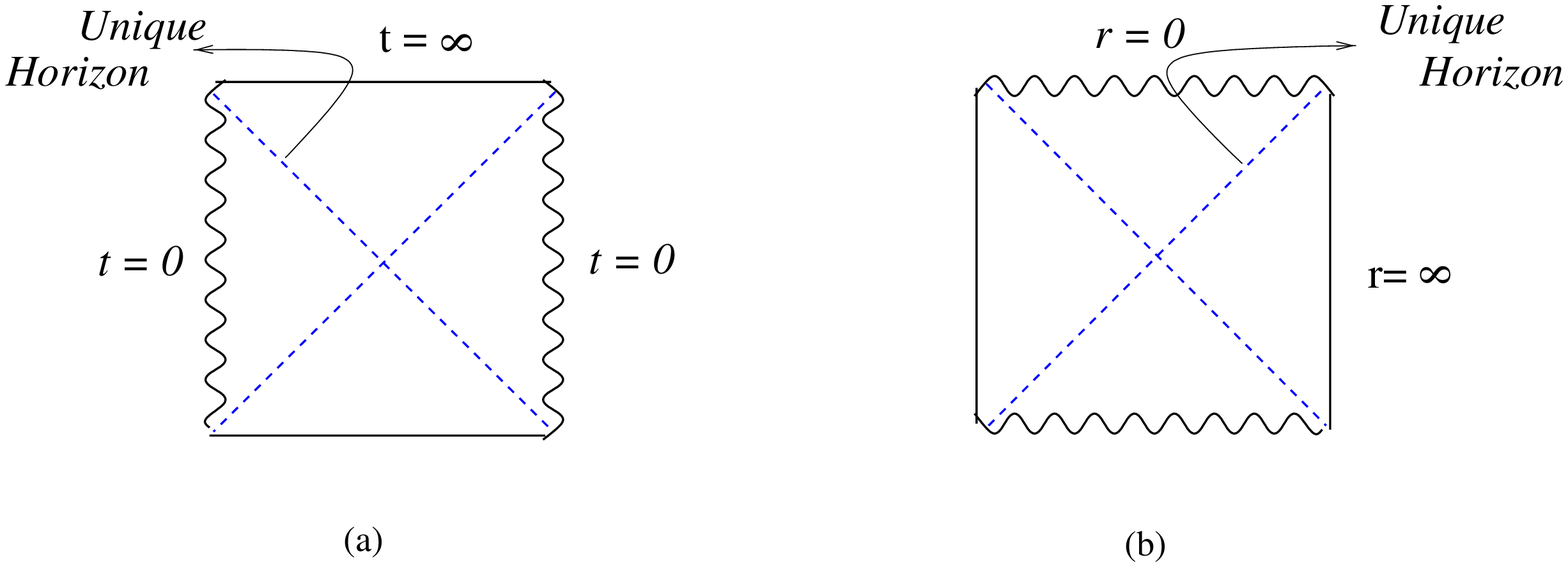, width=.9\textwidth}%
\caption{Penrose diagram for $(a)$ a de~Sitter--S-brane-like
geometry and $(b)$ a Schwarzschild--anti-de~Sitter-like geometry.
The scalar potential prevents the solutions from being
asymptotically flat. \label{noflat}}}

The number of horizons similarly depends on the sign of the scalar
potential. For $\Lambda < 0$, the static solutions can have (depending
on the values of $M$ and $Q$) zero, one or two horizons, respectively
corresponding to the causal structure of the AdS, Schwarzschild-AdS,
or Reissner-Nordstr\"om-AdS geometries. For $\Lambda > 0$, on the
other hand, the time-dependent solutions can have at most one
cosmological horizon, and always have a naked singularity at the
origin (figure~\ref{noflat}). Consequently, these time-dependent
geometries violate the Penrose cosmic censorship hypothesis, which
might give one pause regarding their stability against gravitational
perturbations.

\FIGURE{\epsfig{file=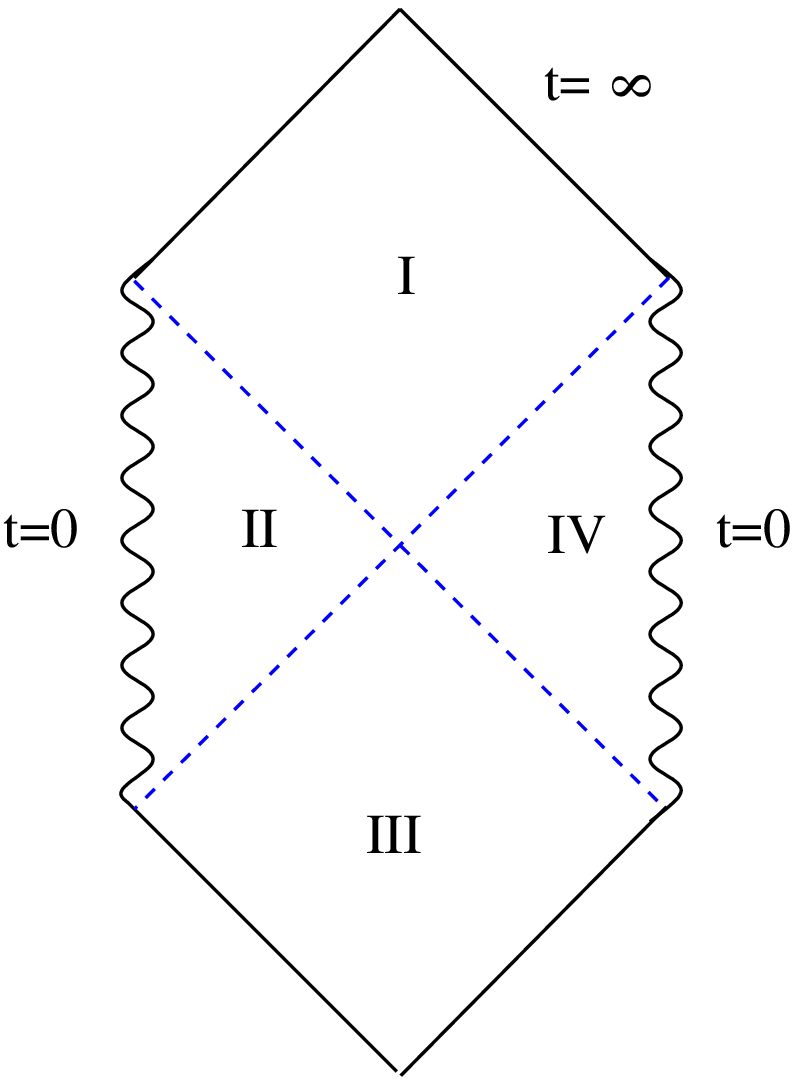, width=.33\textwidth}%
\caption{Penrose diagram for an S-brane.\label{sbrpen}}}

The situation is different if $\beta\rho^2 > \alpha\,(n+1)$. In this
case the geometry is always asymptotically flat and asymptotically
time-dependent, for most choices of $\Lambda$. The causal structure
for this geometry is illustrated by the Penrose diagram of
figure~\ref{sbrpen}. All of the solutions in this class also share a
naked singularity at $r = 0$.

For $\Lambda \to 0$, the solutions in this class reduce directly to
the solutions found in~\cite{gqtz,bqrtz} for $k = q = 0$. The
solutions are therefore asymptotically flat in this limit, and their
Penrose diagram is the same as for an S-brane, see
figure~\ref{sbrpen}. This shows there is a smooth connection between
the S-brane solutions with and without a Liouville potential, and our
new solutions potentially acquire an interpretation as the
supergravity description of a decay of a non-BPS brane, or a $D-\bar
D$ system to the same degree that this is true in the $\Lambda = 0$
case.

{\sloppy
For sufficiently negative $\Lambda$ this example shows explicitly how,
for proper choice of the conformal couplings, the addition of a
negative potential converts the solution into an asymptotically
static, rather than time-dependent, one. The Penrose diagram in the
static case becomes figure~\ref{adsrn}.

}

\FIGURE[b]{\epsfig{file=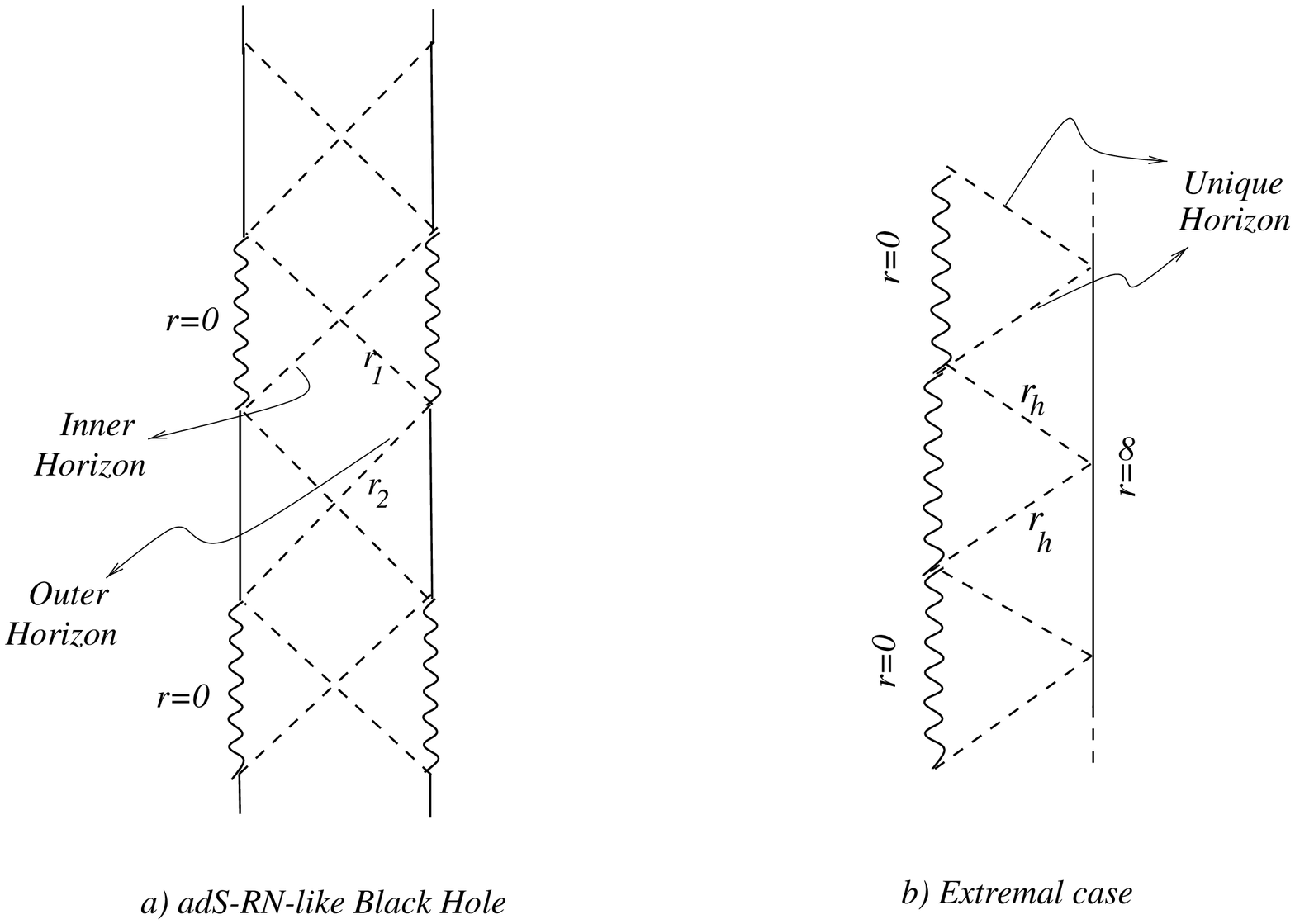, width=.9\textwidth}%
\caption{Penrose diagram for some of our solutions. The causal
  structure is like that of an anti-de Sitter-Reissner-Nordstr\"om
  Black hole $(a)$, or its extremal limit $(b)$.\label{adsrn}}}

{\sloppy
Let us now discuss in detail a specific four-dimensional example,
where the various global properties just discussed can be made
explicit in a transparent way.

}

\paragraph{A four dimensional  example.} 
Consider in detail the 4D case corresponding to $n=2$ and $q=0$,
corresponding to a point source with three transverse dimensions.  We
therefore also have $\M^2 = 2$ and $N = 1$. Since $q=0$, we are free
to choose $Q$ arbitrarily (unlike the requirement $Q = 0$ which follow
if $q \ne 0$). If we use the standard normalizations, $\alpha=1$ and
$\beta=1/2$, $\eta=1/4$, then with the Class I consistency condition
$\sigma = - \lambda$ we may take $\lambda$ and $\Lambda$ as the only
free parameters in the lagrangian.

In this case the metric is
\be
ds^2 =  -h(r) dt^2 +\frac{dr^2}{g(r)} + r^2 dx_{2}^2\,,
\ee
with
\be\label{cs1}
h(r)=-2 M r^{\frac{\rho^{2}}{2}-1}+\frac{Q^{2}}{4(1
    +\frac{\rho^{2}}{2})}\frac{1}{r^{2}}
-\frac{\Lambda r^{2}}{(6-\rho^{2})}
\ee
and
\be g(r)=h(r) r^{-\rho^{2}} \, , 
\ee
where the parameter $\rho$ is given by $\rho = \lambda \sqrt2$.  The
dilaton and gauge fields are similarly given
by~(\ref{dilaton}),~(\ref{qform}) with the appropriate choices for the
various parameters. We suppose, for the time being, that the
parameters $\Lambda$, $M$, and $Q$ are all nonzero.

Consider first the case where the geometry is asymptotically static,
corresponding to $\Lambda < 0$ and $\rho^{2} < 6$. In this case we
have a space-time with the same conformal structure as the
Reissner-Nordstr\"om geometry. This configuration admits an extremal
limit in which the event and Cauchy horizons coincide.  Since the
Hawking temperature for the extremal configuration vanishes, one might
ask whether the extremal geometry can be supersymmetric and stable. In
the next subsections we answer the supersymmetric question by
embedding the solution into particular gauged supergravities for which
the supersymmetry transformations are known, and find the extremal
configurations break all of the supersymmetries. As such it need not
be stable either, although we have not performed a detailed stability
analysis.  There are choices for which the solution~(\ref{cs1}) can
leave some supersymmetries unbroken, however. For instance this occurs
for the domain wall configuration obtained by sending both $M$ and $Q$
to zero.

Let us now specialize to the case $\Lambda >0$. It is then easy to see
that the Ricci scalar, ${\mathcal R}$, for this geometry vanishes at
large $r$ whenever $\rho^{2} > 0$. The Ricci tensor, on the other
hand, does not always similarly vanish asymptotically, and we have
three different cases:

\begin{itemize}
\item When $0\le\rho^{2} \le 2$ the Ricci tensor does not vanish at
  infinity, and so the geometry is asymptotically neither flat nor a
  vacuum spacetime ($R_{\mu\nu} = 0$). The geometry is then
  asymptotically time dependent, with a Cauchy horizon and a pair of
  naked singularities and a Penrose diagram given by
  figure~\ref{noflat}$a$.
\item For $2<\rho^{2}<6$ the Ricci tensor vanishes at infinity, but
  the geometry is nevertheless not asymptotically flat since the
  Riemann tensor is nonzero at infinity. The spacetime's causal
  structure is much as in the previous case.
\item When $\rho^{2}>6$, the geometry is asymptotically flat and
  time-dependent, with the Penrose diagram of
  figure~\ref{sbrpen}. Notice that in this case the time-dependence of
  the metric is \emph{not} due to the scalar potential, but instead
  arises from the choice of the sign of integration constants. In
  particular, the geometry is time dependent for positive $M$, and
  also in the case $\Lambda<0$.
\end{itemize}

The interesting feature of this example is that it brings out how the
asymptotic properties change with the choice of $\rho$. It also shows
that the geometries can be asymptotically time-dependent even when the
scalar potential is negative (corresponding to AdS-like curvatures).
These solutions share the Penrose diagram --- figure~\ref{sbrpen} --
of the S-brane geometries of ref.~\cite{bqrtz}, and in this way
generalize these to negative scalar potentials. The geometry shares
the naked singularities having negative tension of the S-brane
configurations, and we expect many of the arguments developed in that
paper to go over to the present case in whole cloth.

\paragraph{Class II.}
Solutions belonging to this class are similar to --- but not identical
with --- the ones of the previous class. Choosing $\Lambda < 0$ we
again find static solutions with at most two horizons (as for
figure~\ref{adsrn}). When $\Lambda > 0$, on the other hand, we have
asymptotically time-dependent solutions with at most one horizon and a
naked singularity at the origin. It is also not possible to obtain an
asymptotically flat configuration for either sign of $\Lambda$ or for
any choice of the conformal coupling. When $\Lambda \to 0$, these
solutions again reduce to the asymptotically-flat S-brane geometries
of~\cite{bqrtz}.

\paragraph{A four dimensional example.}
Specializing to 4 dimensions we choose $n=2$ and $q = 0$, and so $\M^2
= 2$ and $N=1$. Since $q = 0$, $Q$ can remain arbitrary.  With the
conventional choices $\alpha=1$, $\beta=1/2$ and $\eta=1/4$, the
consistency condition $\alpha \sigma \lambda \M^2 + 4 \beta = 0$
requires $\sigma = -1/\lambda$, leaving as free parameters $\lambda$,
$\Lambda$, and the integration constants $M$ and $Q$. The constant
$\rho$ is then given by $\rho = \sqrt2 /\lambda$.

With these choices the metric becomes
\be 
ds^2 =  -h(r) dt^2 +\frac{dr^2}{g(r)} + r^2 dx_{2}^2\,, 
\ee
with
\be
\label{exe3} h(r)=-2 M r^{\frac{\rho^{2}}{2}-1}-\frac{ \Lambda
r^{\rho^{2}}}{ \rho^{2} \left[ 1+ \frac{\rho^{2}}{2} \right]}
+\frac{Q^{2}}{4\left[ 1+ \frac{\rho^{2}}{2} \right]r^{2}} \,, 
\ee
and
\be 
g(r)=h(r) r^{-\rho^{2}}\,. 
\ee
As before the expressions for the scalar and the 2-form field strength
are easily obtained from formulae~(\ref{dilaton}),~(\ref{qform}), with
the appropriate substitutions of parameters. From \linebreak
eq.~(\ref{exe3}) it is clear that the term proportional to $\Lambda$
is dominant for large $r$ for all $\rho^2 > 0$. This means that for
$\Lambda < 0$ we have a static solution for large $r$, while for
$\Lambda > 0$ it is time-dependent in this limit.

\paragraph{Class III.}
This class contains the most interesting new time-dependent
configurations. Here, as in Class I, the asymptotic properties of the
geometry depend on the choice of the parameter $\rho$, and so
ultimately also on the conformal couplings.

Consider first $\beta\rho^2 < \alpha$. For $\Lambda < 0$ this gives
static configurations with at most one horizon --- which are neither
asymptotically flat nor asymptotically AdS --- having the same
conformal structure as Schwarzschild-AdS black branes. For $\Lambda >
0$ by contrast the solution is asymptotically time dependent (but not
asymptotically flat) and can have up to two horizons, implying the
same conformal structure as for a Schwarzschild-de Sitter black hole
(see figure~\ref{scds}).  Interestingly, these properties imply an
asymptotically time-dependent configuration, but \emph{without} naked
singularities.

\FIGURE[t]{\epsfig{file=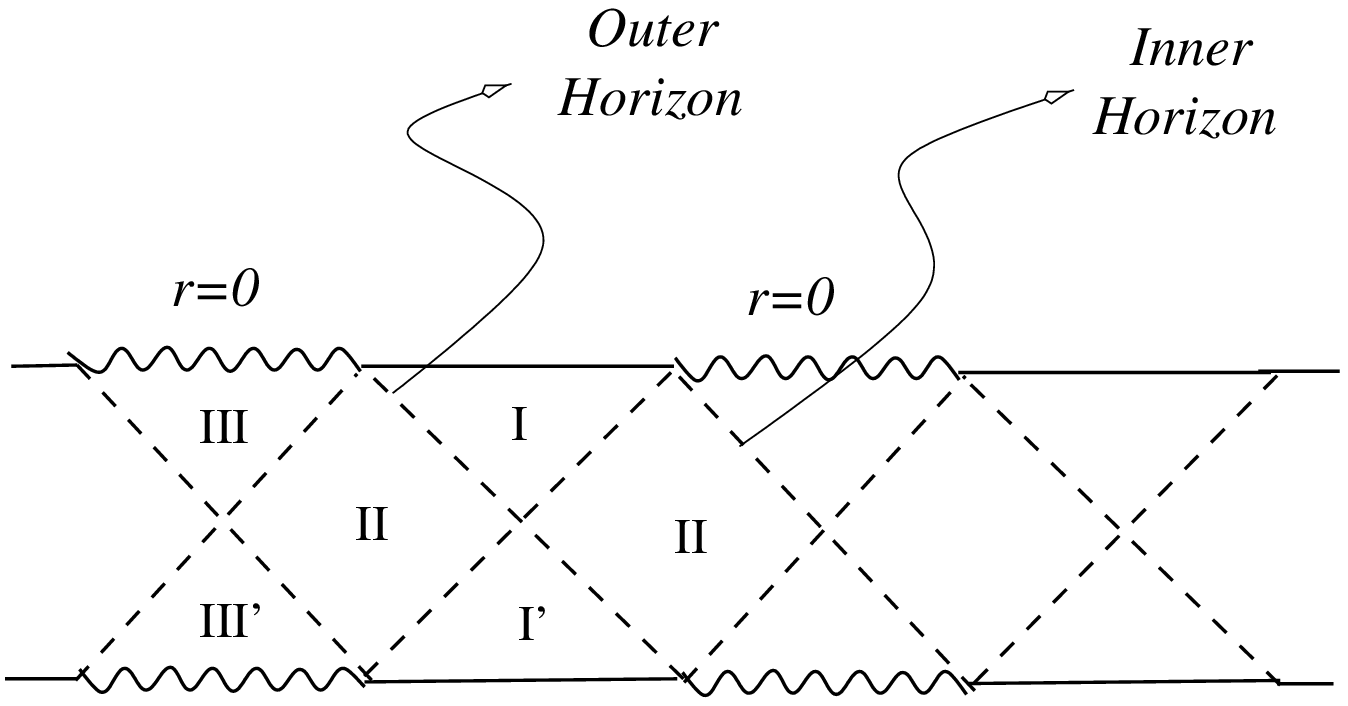, width=.6\textwidth}%
\caption{The Penrose diagram for $\Lambda > 0$ Class III solutions,
  which resembles the same for a Schwarzschild-de Sitter black hole
  geometry.\label{scds}}}

If, however, $\beta\rho^2 > \alpha$ then the choice $\Lambda > 0$
instead gives static configurations having at most one horizon.
Choosing $\Lambda < 0$ also gives solutions are which are static,
although for these one finds at most two horizons if $\beta\rho^2 >
\alpha (n+1)$, or only one horizon if $\alpha < \beta\rho^2 < \alpha
(n+1)$.

These solutions do not reduce to Ricci-flat space-times in the limit
$\Lambda \to 0$. This limit instead gives a class of non-standard
black brane solutions, which are included among the general class of
solutions found in~\cite{cina} (and were interpreted in~\cite{oz} in
terms of unstable branes).

\paragraph{A four dimensional example.}
We now specialize to the 4-dimensions with the choice $n=2$ and $q =
0$, implying $\M^2 = 2$ and $N = 1$. Just as was true for Class II,
the usual choices $\alpha=1$, $\beta=1/2$ and $\eta=1/4$ in this case
imply that the consistency condition $\alpha \sigma \lambda (n+q) = 4
(n-1) \beta$ reduces to $\sigma = 1/\lambda$.  Unlike for Class II,
however, the expression for $\rho$ may be solved, giving $\rho =
\lambda \sqrt2$.

The metric becomes:
\be 
ds^2 =  -h(r) dt^2 +\frac{dr^2}{g(r)} + r^2 dx_{2}^2\,, 
\ee
with
\be
\label{exe2} 
h(r)=-2 M
r^{\frac{\rho^{2}}{2}-1}-\frac{ \Lambda \, r^{2}}{(6-
\rho^{2})}+\frac{Q^{2} \, r^{\rho^{2}}}{\rho^{2}(2+\rho^{2})} 
\ee
and
\be 
g(r)=h(r) r^{-\rho^{2}} \,, 
\ee
and the expressions for the scalar and the form are found from
formulae~(\ref{dilaton}),~(\ref{qform}) specialized to the relevant
parameter values.

We now focus on the asymptotically time-dependent geometries since
these are qualitatively new, and since the asymptotically-static
solutions are similar to those found in the previous classes. To this
end consider the case $\rho^{2} < 2$ and $\Lambda > 0$, for which the
metric is asymptotically time dependent with horizons which cover the
geometry's space-like singularities.

With these choices the Ricci scalar vanishes at infinity (although
the Ricci tensor does not so long as $\rho \neq 0$), and so the
metric is neither asymptotically flat nor asymptotically a vacuum
spacetime. Neither is it asymptotically de Sitter, even though
$\Lambda > 0$, but it is instead a kind of interpolation between a
de Sitter geometry and an S-brane configuration.

The Penrose diagram for the geometry resembles that of a de
Sitter-Schwarzschild space-time, as in figure~\ref{scds}, for which
the hiding of the singularities by the horizons is clear.  This
satisfies the cosmic censorship conjecture, and one might hope it to
be stable against the perturbations which were argued~\cite{bmqtz} to
potentially destabilize the $\Lambda = 0$ S-brane
configuration.\footnote{See, however, ref.~\cite{cc} who argue that
  these geometries may be stable if the boundary conditions are chosen
  appropriately.}

The asymptotically time-dependent regions of the geometries are those
labelled by I and I' in figure~\ref{scds}. In region I' the spatial
geometry contracts as one moves into the future, while in region I the
spatial geometry expands. As the figure makes clear, it is possible to
pass smoothly from the contracting to the expanding region without
ever meeting the singularities.  Remarkably, neither does one pass
through a region of matter satisfying an unphysical equation of state,
such as is typically required to produce a bouncing FRW universe. We
believe it to be worthwhile trying to extend these geometries (perhaps
to higher dimensions) to construct more realistic bouncing cosmologies
without naked singularities or unphysical matter.

\section{Gauged and massive supergravities}\label{part3}

The exponential Liouville potential we consider above has many
practical applications because it frequently arises in the bosonic
sector of gauged supergravities. However the solutions we construct
require specific relations to hold amongst the various conformal
couplings of the general action,~(\ref{generalaction}), and it must be
checked that these are consistent with supersymmetry before
interpreting our geometries as solving the field equations of any
particular supergravity.

In this subsection we verify that many extended supergravities do
satisfy the required conditions, and by so doing find numerous
solutions to specific supergravity models. We find that many of these
describe the field configurations due to extended objects in the
corresponding supergravity.

Some of the supergravities we consider have well-established string
pedigrees inasmuch as they can be obtained as consistent truncations
of ten-dimensional type-IIA and IIB supergravities, which themselves
arise within the low-energy limit of string theory. This allows the
corresponding solutions to be lifted to ten dimensions, and so they
themselves may be interpreted as {\it bona fide} low-energy string
configurations. It also allows them to be related to each other and to
new solutions using the many symmetries of string theory such as
\emph{T-duality} and {\it S-duality}.

We organize our discussion in the order of decreasing dimension,
considering the 10-, 8-, 7-, 6-, 5- and 4-dimensional cases. Among
these we consider examples of massive, compact-gauged and
non-compact-gauged supergravities. Of these the study of compact
gauged supergravities is particularly interesting due to its relation
with extensions of the AdS-CFT correspondence. On the other hand it is
the non-compact gauged supergravities which are most likely to have
cosmological applications.

\subsection{Massive supergravity in 10 dimensions}

Romans~\cite{romans} has shown how to construct a ten-dimensional
supergravity theory which has an exponential scalar potential for the
dilaton, and we take this as our first example. The solution we obtain
in this case was earlier obtained in refs.~\cite{perry,berg}.

The bosonic fields of the theory comprise the metric, a scalar, and
2-form, 3-form and 4-form field strengths, $F_2 = \exd A_1$, $F_3 =
\exd A_3$ and $F_4 = \exd A_3$. The equations of motion for all of the
fields is trivially satisfied if we set all of their field strengths
to zero, leaving only the dilaton and the metric. The relevant action
for these fields is~\cite{berg}:
\be
    \label{romansaction}
    S= \int d^{10}x \sqrt{|g|} \left[  R - \frac{1}{2} (\partial
    \phi)^2 -\frac{1}{2} m^{2} {e}^{5\phi/2}\right]
\ee
which is a special form of eq.~\pref{generalaction} obtained by
choosing $\alpha = 1$, $\beta = \frac12$, $\Lambda = \frac12 \,
m^2$ and $\lambda = - 5/2$. Notice we need not specify
either $\eta$ or $\sigma$ since we may turn off the gauge form
field simply by choosing $Q=0$.

To obtain a solution we choose $q=0$, and from the relation $n + q + 2
= 10$ we have immediately $n=8$. We find a solution in Class I, for
$k=0$, with $\M^2 = 8$ and $\rho = 2\sqrt2 \lambda = -5
\sqrt{2}$. This leads to the following solution (which is the same as
discussed in~\cite{perry})
\bea
d s_{10}^{2} &=& -h(r) d t^{2}+\frac{r^{50}}{h(r)} d r^{2}+ r^{2}d x_{0,8}^{2}
\nonumber\\ 
\phi(r) &=& - 20 \ln r 
\eea
where
\be
h(r)=\frac{m^{2}}{256} r^{2}-2M r^{18}\,.
\ee

For $M>0$ this solution is time-dependent for large $r$, and has a
horizon at $r^{16} = m^2/(512 M)$. It is interesting to note that we
would obtain an asymptotically time-dependent configuration even if
the scalar potential were to be negative, due to the dominance at
large $r$ of the term proportional to $M$. The geometry in this case
is asymptotically flat, and the Penrose diagram is the same as for an
S-brane solution, shown in figure~\ref{sbrpen}.

The case $M=0$ corresponds instead to a static space-time, and a
particularly interesting one at that since a straightforward check
reveals that it is supersymmetric since it admits a Killing spinor
preserving half of the supersymmetries of the action. In this case,
the metric becomes (after a constant rescaling of the eight internal
spatial coordinates):
\be 
d s_{10}^{2} = h(r)\,\eta_{\mu \nu} d x^{\mu}d x^{\nu}
+\frac{r^{50}}{h(r)} d r^{2}\,. 
\ee
In this case the solution belongs to the family discussed in
section~\ref{section4} of~\cite{berg}, the so-called ``domain wall''
solution. Notice also that original $\SO(8) \times \SO(1,1)$ symmetry
is promoted in this case to $\SO(1,8) \times {\mathbb R}$, where
${\mathbb R}$ denotes translation symmetry in the $r$ direction.

\subsection{Gauged supergravity in 8 dimensions}\label{section4}

We next consider Salam-Sezgin gauged supergravity in eight
dimensions~\cite{Salam:1984ft}. In general, the bosonic part of the
action consists of the metric, a dilatonic scalar, five scalars
parameterizing the coset $\SL(3, \mathbb R)/\SO(3)$ (and so consisting
of a $3 \times 3$ unimodular matrix, $L$), an $\SU(2)$ gauge potential
and a three-form potential $C_{\it 3}$. We further restrict ourselves
to a reduced bosonic system where the $\SU(2)$ gauge fields vanish,
and we take $L$ to be a constant diagonal matrix.

The action for the remaining bosonic fields can be written
\be 
S=\int d^{8}x \sqrt{|g|}
\left[\frac{1}{4}R-\frac{1}{2}(\partial \phi)^{2} -\frac{1}{48}
e^{2 \phi} G_{\mu \nu \lambda \sigma} G^{\mu \nu \lambda \sigma}
+{\rm g}^{2} e^{-2 \phi}  \right] \, ,
\ee
and so we read off $\alpha = 1/4$, $\beta = \eta = 1/2$,
$\sigma = -2$, $\Lambda = -{\rm g}^2$ and $\lambda = 2$. The 3-form
potential couples to $q = 2$ branes, so in $n + q + 2 = 8$ dimensions
we have $n = 4$.

In this case we can find solutions in class I, since this model
satisfies the appropriate consistency condition $\sigma = -
\lambda$. Consequently $k = 0$ and since $q \ne 0$ the solution also
has $c = 1$ and describes fields for 2-branes carrying vanishing
charge, $Q = 0$. With these values we find $\M^2 = 6$, $N = 1$ and
$\rho$ is predicted to be $\rho = \sqrt{3/2}$, and the solution takes
the following form:
\bea 
ds_{8}^{2}&=&-h(r)d
t^{2}+r^{2}dx_{0,4}^{2}+\frac{r^{6}}{h(r)} dr^{2}
+r^{2}dy_{2}^{2} \,,\\
 \phi(r) &=& 3 \ln r\,,\\
 h(r)&=&-\frac{2M}{r^{2}}+ \frac{{\rm g}^{2} r^{2}}{6}\,.
\eea

This geometry correspond to an uncharged black 2-brane, and has at
most one horizon. The causal structure of the geometry resembles the
AdS-Schwarzschild black hole. These field configurations turn out to
be supersymmetric only when $M = Q = 0$.

\paragraph{Uplifting procedure.}
\looseness=-1 Because this supergravity can be obtained by
consistently truncating a higher-dimensional theory, this solution
also directly gives a solution to the higher-dimensional field
equations. We now briefly describe the resulting uplift to 11
dimensions, following the procedure spelled out
in~\cite{Salam:1984ft}. The corresponding 11-dimensional metric and
4-form are,
\be
\label{elevend}
ds_{11}^2= e^{-2\phi/3}ds_{8}^2  + e^{4\phi/3} (d\Omega_3)^2\,,
\ee
or,
\be ds_{11}^2=   -\frac{h(r)}{r^2}dt^2 + dx_{4}^2 + dy_{2}^2 +
\frac{r^4}{h(r)}dr^2 + r^4(d\Omega_3)^2\,. 
\ee
The form of this 11-dimensional solution makes it clear why the choice
$M = Q = 0$ is supersymmetric, because in this case the preceding
metric describes flat 11-dimensional space.

\subsection{Gauged supergravity in 7 dimensions}\label{7dtv}

In this subsection we present two kinds of solutions to the
7-dimensional $\mathcal N=2$ gauged supergravity with $\SU(2)$ gauge
symmetry studied in~\cite{townew}. The bosonic part of the theory
contains the metric tensor, a scalar $\phi$, $\SU(2)$ gauge fields,
$A^{I}_\mu$, with field strengths, $F^{I}_{\mu \nu}$, and a 3-form
gauge potential, $C_{\mu \nu \rho}$, with field strength $G_{\mu\nu
  \lambda\rho}$. The bosonic action for the theory is~\cite{townew}:
\be
{\rm e}^{-1} \mathcal{L}  =  R
-\frac{1}{2}(\partial \phi)^{2}-\frac{1}{4}
e^{\frac{4}{\sqrt{10}}\, \phi} G_{\mu\nu\lambda\rho}
 G^{\mu\nu\lambda\rho} -\frac{1}{4}
e^{-\frac{2}{\sqrt{10}} \phi} F^{I}_{\mu\nu}
 F^{I \,\mu\nu }
+ 4 {\rm g}^{2} e^{\frac{2}{\sqrt{10}} \phi}
+\frac{1}{4} {\rm e}^{-1} F^{I}_\emph{2} \wedge F^{I}_\emph{2} \wedge
 A_\emph{3} \,,
\label{ac7dtv} 
\ee
where `${\rm e}$' corresponds to the determinant of the 7-bein, and
${\rm g}$ is the gauge coupling of the $\SU(2)$ gauge group. {}From
this action we read off $\alpha = 1$, $\beta = 1/2$, $\Lambda = -
4 {\rm g}^2$ and $\lambda = -2/\sqrt{10}$.

We now consider the following two kinds of solution to this theory,
which differ by whether it is the 2-form, $F^I_\emph{2}$, or the
4-form, $G_\emph{4}$, which is nonzero. In either of these cases a
great simplification is the vanishing of the Chern-Simon terms vanish.

\subsubsection{Solutions with excited $F^{I}_\emph{2}$}\label{sim7sol}

The solution for to nonzero $F_\emph{2}^I$ corresponds to the field
sourced by a point-like object, $q = 0$, and so $n + q + 2 = 7$
implies $n = 5$. This particular tensor field satisfies $\eta =
\frac12$ and $\sigma = 2/\sqrt{10} = - \lambda$.

Since $\sigma = - \lambda$ we look for solutions belonging to Class
I. For these $k = 0$, $Q$ is not constrained because $q = 0$. Also
$\M^2 = 5$, $N = 1$ and the parameter $\rho = \lambda \M = -
\sqrt{2}$. The corresponding metric takes the form
\be
\label{sevenmet}
d s_{7}^{2} = -h(r) d t^{2} + r^{2}d x_{0,5}^{2}
+\frac{r^{2}}{h(r)} d r^{2}  \,,
\label{7dtnF} 
\ee
with metric coefficients, scalar and 2-form field strength given
by
\bea h(r) &=& -\frac{2M}{r^{3}}+\frac{4 {\rm
g}^{2}}{25}r^{2}+\frac{Q^{2}}{50 \, r^8} \\
\phi(r) &=& -\sqrt{10} \ln r \\
F^{\rm tr} &=& \frac{Q}{r^{8}}\epsilon^{\rm tr} \,.
\eea

This solution is static and corresponds to a charged black hole in
seven dimensions. It can have up to  two horizons corresponding to
the two zeros of $h(r)$, which can be explicitly found as
\be
h(r)= \frac{4\g^2}{25\,r^8} \left(-\frac{50M}{4\g^2} + r^{10}
                          + \frac{Q^2}{8\g^2}\right) =
    \frac{4\g^2}{25\,r^8} (X-X_1)(X - X_2)\,,
\ee
where  $X= r^5$ and
\be
X_{1,2} = \frac{25M}{4\g^2} \pm \frac{1}{2\g}
\sqrt{\frac{(25M)^2}{4\g^2} -\frac{Q^2}{2}}\,.
\ee
From here is clear that there would be an extremal solution when $Q=
\frac{25M}{\sqrt{2}\g}$. One can further check that the above field
configuration becomes supersymmetric if the parameters $M$ and $Q$ are
taken to zero. Notice that this supersymmetric configuration is
\emph{not} the extremal configuration, for which the two horizons
coincide at a double zero of $h(r)$.

\paragraph{Uplifting procedure.}
Following~\cite{chamse} this solution can be oxidized on a three
sphere $S^{3}$ to give a solution to ten dimensional IIB
supergravity. This 10D theory contains a graviton, a scalar field, and
the NSNS 3-form among other fields, and has a ten dimensional action
(in the Einstein frame) given by
\be 
S_{10}=\int d^{10}x \sqrt{|g|} \left[ \frac{1}{4} R-\frac{1}{2} (\partial
\phi)^{2} -\frac{1}{12} \, e^{-2 \phi} H_{\mu \nu \lambda}H^{\mu
\nu \lambda}\right]\,. \ee

We perform some conventional rescalings to convert to the
conventions of~\cite{chamse}, after which we have a ten
dimensional configuration given by
\bea{\label{7d10d}}
ds_{10}^2 &=& \left(\frac{2}{r}\right)^{3/4}\left[ ds_7^2 \right] +
\left(\frac{r}{2}\right)^{5/4} \left[ d\theta^2 + d\psi^2 + d\varphi^2 +
(d\psi + \cos\theta d\varphi -\sfrac{Q}{5r^5}dt)^2\right],
\nonumber\\ 
\phi &=& -\frac{5}{4}\,\log\frac{r}{2}  \,,
\nonumber\\ 
H_3 &=& -\frac{ Q }{r^{6}}\, \exd r\wedge \exd t\wedge(\exd\psi +
\cos\theta \exd\varphi) - \frac{{\rm g}}{\sqrt{2}} \sin\theta \,
\exd\theta \wedge \exd\varphi\wedge \exd\psi        \,.
\eea
where $d s_{7}^2$ corresponds to the solution given in
eq.~(\ref{sevenmet}). This uplifted 10-dimensional solution describes
NS-5 branes intersecting with fundamental strings in the time
direction.

\paragraph{S-duality.}
For a later comparison with the uplifted solution of 6-dimensional
supergravity, it is convenient to rewrite this 10-dimensional solution
by performing a change of variables and a new re-scaling of the
parameters. For the same reason, it is also useful to perform in this
subsubsection an \emph{S-duality} transformation of this solution.

As a first step to this end, let us make the manipulation of the
angular variables of the three sphere simpler by introducing the
following left-invariant 1-forms of $\SU(2)$:
\bea\label{sigmas}
    \sigma_1 &=& \cos\psi \, \exd\theta + \sin\psi \sin\theta \,
    \exd\varphi \,, 
\nonumber\\ 
    \sigma_2 &=& \sin\psi \, \exd\theta - \cos\psi \sin\theta
    \, \exd\varphi \,, 
\nonumber\\ 
    \sigma_3 &=& \exd\psi + \cos\theta \, \exd\varphi \,,
\eea
and
\be 
h_{3}=\sigma_{3}- \frac{Q}{5}\frac{1}{r^5}\, \exd t\,. 
\ee
Next, we perform the following change of variables
\be
\begin{array}[b]{rclrclrclrcl}
\label{changeseven}
\displaystyle\frac{r}{2}&=&\rho^{\frac{4}{5}}\,,
&
t &=&\displaystyle \frac{5}{32} \,\tilde{t}\,, 
&
\exd x_{4} &=&\displaystyle\frac{1}{2\sqrt{2}}
    \, \exd \tilde{x}_{4}\hskip.3cm\,,
\qquad &
\exd x_{5}&=&\displaystyle\frac{1}{2} \, \exd Z 
\\
    {\rm g}&=&\sqrt{2} \, \tilde{{\rm g}}\,,
\qquad &
Q&=& \sqrt{2} \, 2^{7} \,
    \tilde{Q} \,,
\qquad &
    \sigma_{i}&=&\displaystyle\frac{1}{\tilde{{\rm g}}}\,
    \tilde{\sigma}_{i} \,. &&&
\end{array}
\ee
It is straightforward to check that the 10-dimensional
solution~(\ref{7d10d}) becomes, after these changes
\bea\label{7d10dchange}
    d \tilde{s}_{10}^{2} &=& \frac{1}{2}\, \rho^{-1} \,\left[d
    \tilde{s}_{6}^2 \right]
    +\frac{\rho}{\tilde{{\rm g}}^{2}}\left[\tilde\sigma_{1}^{2}
                     +\tilde\sigma_{2}^{2}+\left(\tilde\sigma_{3}
     -\frac{\tilde{{\rm g}}\tilde{Q}}{4 \sqrt{2}}\frac{1}{\rho^{4}}
    d \tilde{t} \right)^{2} \right]+\rho\, d Z^{2} \,, 
\nn \\
    \phi &=& - \ln \rho\,,  
\nn \\
    H_\emph{3} &=& -\frac{1}{\tilde{{\rm g}}^{2}} \tilde{\sigma}_{1} \wedge
    \tilde{\sigma}_{2} \wedge \tilde{h}_{3} +\frac{\tilde{Q}}{\sqrt{2}
    \, \tilde{{\rm g}} \, \rho^{5}} \exd\tilde{t} \wedge \exd \rho \wedge
    \tilde{h}_{3} \, ,
\eea
where we define
\be
\label{seventosix}
    d \tilde{s}_{6}^2=-\tilde{h}(\rho)\,
    d\,\tilde{t}^{2}+\frac{\rho^{2}}{\tilde{h}(\rho)} d\, \rho^{2}+ \rho^{2}\,
    d \tilde{x}_{0,4}^{2}
\ee
and, after re-scaling $M$,
\be
\label{hseventosix}
    \tilde{h}=-\frac{2\tilde{M}}{\rho^{2}}+\frac{\tilde{g}^{2}}{32}\rho^2
    +\frac{\tilde{Q}^{2}}{8}
    \frac{1}{\rho^{6}} \,.
\ee

We now transform the solution from the Einstein to the string
frame (and denote all string-frame fields with a bar). This leads
to
\bea
\label{7d10dstring} 
d \bar{s}_{10}^{2} &=& \frac{1}{2}\,
\rho^{-2} \,\left[d \tilde{s}_{6}^2 \right]
+\frac{1}{\tilde{{\rm g}}^{2}}\left[\tilde{\sigma}_{1}^{2}
             +\tilde{\sigma}_{2}^{2}
+\left(\tilde{\sigma}_{3} -\frac{\tilde{{\rm g}}\tilde{Q}}{4
\sqrt{2}}\frac{1}{\rho^{4}}
d \tilde{t} \right)^{2} \right]+ d Z^{2}\,, 
\nn \\
\bar{\phi} &=& -2 \ln \rho\,,  
\nonumber\\ 
\bar{H_\emph{3}}  &=& H_\emph{3}  \, .
\eea

We have a solution to 10-dimensional IIB supergravity with a
nontrivial NSNS field. If we perform an S-duality transformation
to this solution we again obtain a solution to type-IIB theory but
with a nontrivial RR 3-form, $F_\emph{3}$. The S-duality transformation
acts only on the metric and on the dilaton, leaving invariant the
three form. In this way we are led to the following configuration,
which is S-dual to the one derived above
\bea
\label{7d10dsdual}
d \bar{s}_{10}^{2} &=& \frac{1}{2}\, \,\left[d
\tilde{s}_{6}^2 \right]
+\frac{\rho^{2} }{\tilde{{\rm g}}^{2}}\left[\tilde{\sigma}_{1}^{2}+
\tilde{\sigma}_{2}^{2}
+ \left(\tilde{\sigma}_{3}-\frac{\tilde{{\rm g}}\tilde{Q}}
     {4 \sqrt{2}}\frac{1}{\rho^{4}}
d \tilde{t} \right)^{2} \right]+ \rho^{2} \,d Z^{2}\,,
\nn  \\
\bar{\phi} &=& 2 \ln \rho\,,  
\nonumber\\ 
F_\emph{3} &=& H_\emph{3}\,.
\eea
This form is particularly convenient for making the comparison with
our later uplifted solution to 6-dimensional supergravity.

\subsubsection{Solution with excited $G_\emph{4}$}

We can also apply our ansatz to obtain a solution to 7-dimensional
supergravity which sources the 3-form potential, $A_\emph{3}$, for
which we have $q = 2$ and so $n = 5 - q = 3$. The 3-form couplings in
the lagrangian are $\eta = 6$ and $\sigma = - 4 /\sqrt{10} = 2
\lambda$.

These couplings allow solutions belonging to Class III, and so for
which $c = 1$, $\M^2 = 5$, $N = 1$, and $\rho = \lambda \M =
-1/\sqrt2$. Formulae~(\ref{class3}) allow two possible curvatures for
the $n=3$ dimensional symmetric subspace. It is either spherical
($k=1$) and the brane charge is $Q^{2}=1/3$, or it is flat ($k=0$)
with brane charge $Q=0$.

For the spherical case, the metric takes the form
\be 
d s_{7}^{2} = -h(r) d t^{2} + r^{2}d x_{1,3}^{2}
+\frac{r^{2}}{h(r)} d r^{2} +r^{2}d y^{2}_2\, , 
\label{7dtnG}
\ee
and the metric coefficients, the scalar and the 2-form field are
\bea
h(r) &=& -\frac{2M}{r^{3}}+\left(\frac{4 {\rm g}^{2}}{25}
+\frac{12\,Q^{2}}{25}\right) r^{2} 
\\
\phi(r) &=& -\sqrt{10} \ln r 
\\
F^{try_{1} y_{2}} &=& \frac{Q}{r^{2}} \epsilon^{ t r y_{1} y_{2}}\,. 
\eea

This geometry is static an possesses only one horizon. It describes a
charged black 2-brane in seven dimensions, for which the special case
$M = Q = 0$ is supersymmetric (and coincides with this same limit of
the solution of the previous subsubsection). The uplifting to 10
dimensions can be done straightforwardly by first dualising the
$G_\emph{4}$ field in seven dimensions to transform it into a 3-field,
and then applying the prescription given in ref.~\cite{chamse}.

\subsection{Romans' 6-dimensional gauged supergravity}

The next two subsections consider two kinds of gauged supergravity in
six dimensions. We start here with Romans' $N = 4^g$ 6D supergravity,
which was oxidized to 10 dimensions in ref.~\cite{cpl1}.

Romans'~\cite{romansix} 6-dimensional $N = 4^g$ gauged supergravity is
non-chiral and has ${\mathcal N}=4$ supersymmetries. The bosonic part
of the theory consists of a graviton, three $\SU(2)$ gauge potentials,
$A_{\mu}^{I}$, an abelian gauge potential, ${\mathcal A}_{\mu}$, a
2-form gauge potential, $B_{\mu \nu}$, and a scalar field, $\phi$. Our
starting point is the following consistently reduced version of this
action~\cite{nunez}:
\bea
\label{6daction}
 S&=&\int d^{6}x \sqrt{|g|} \Bigg[\frac{1}{4}R-\frac{1}{2}(\partial \phi)^{2}
  -\frac{1}{4} e^{- \sqrt{2} \,\phi}({\mathcal F}_{\mu\nu}{\mathcal
F}^{\mu\nu} + F_{\mu\nu}^{I}F^{I\,\,\mu\nu})
   -\frac{1}{12}e^{2 \sqrt{2} \phi}G_{\mu\nu\rho}G^{\mu\nu\rho}+ 
\nonumber\\
&&\hphantom{\int d^{6}x \sqrt{|g|} \Bigg[ }\! 
+ \frac{{\rm g}^{2}}{8}e^{\sqrt{2} \phi}
   -\frac{1}{8\sqrt{|g|}}\, \epsilon^{\mu\nu\rho\sigma\tau\kappa}
   B_{\mu\nu} ({\mathcal
  F}_{\rho\sigma}{\mathcal F}_{\tau\kappa} + F^I_{\rho\sigma}
                  F^I_{\tau\kappa}  )\Bigg]\,.
\eea
Here, ${\rm g}$ is the coupling constant of the $\SU(2)$ group, and
$\epsilon^{\mu\nu\rho\sigma\tau\kappa}$ is the usual Levi-Civita
tensor density. $F^I_\emph{2}$ denotes the $\SU(2)$ gauge field
strength, while ${\mathcal F}_\emph{2} = \exd {\mathcal A}_\emph{1}$
and $G_\emph{3} = \exd B_\emph{2}$ are the abelian field strengths for
the abelian potential and the antisymmetric field. The supersymmetry
transformations with these conventions may be found, for instance,
in~\cite{nunez}. This action leads to the choices $\alpha = 1/4$,
$\beta = 1/2$, $\Lambda = - {\rm g}^2/8$ and $\lambda = - \sqrt2$.

We consider in turn two cases which differ in which gauge potential is
excited by the solution. We first take this to be the 2-form
potential, $B_\emph{2}$, and then choose it to be one of the 1-form
potentials, $A^I_\mu$ or ${\mathcal A}_\mu$. In this second case the
solution looks the same for either choice of embedding within the
gauge group, although the supersymmetry of the result can differ.

\subsubsection{Solutions with excited $G_\emph{3}$}\label{6exB}

We first consider a charged string which sources the field $B_{\it
  2}$, and so for which $q=1$, $n = 4 - q = 3$ and the dilaton
couplings are $\eta = 1/2$ and $\sigma = -2 \sqrt2 = 2 \lambda$. These
couplings allow solution belonging to Class III, implying $c = 1$,
$\M^2 = 4$, $N = 1$ and $\rho = - 1/\sqrt2$. The curvature of the $n=
3$ dimensional subspace can be $k = 1$ if $Q^{2}=1$, or it is flat if
$Q^{2}=0$.

For the case $k = 1$ the metric takes the form
\be
d s_{6}^{2} = -h(r) d t^{2} + r^{2}d x_{1,3}^{2}
+\frac{r^{2}}{h(r)} d r^{2} +r^{2}d y^{2} \label{6dfc}
\ee
with metric coefficients, scalar and 2-form field given by
\bea
h(r) &=& -\frac{2M}{r^{2}}+
\frac{{\rm g}^{2}}{32}r^{2}+\frac{Q^{2}}{2}r^{2}\,, 
\\
\phi(r) &=& -\sqrt{2} \ln r \,, 
\\
G^{try} &=& \frac{Q}{r} \epsilon^{try} \,.
\eea

The geometry describes the fields of a charged black string in six
dimensions. In the limit where $M$ and $Q$ vanish, (and so for which
the symmetric 3-dimensional subspace is flat) the solution preserves
half of the supersymmetries of the action.

\paragraph{Uplifting procedure.}
Ref.~\cite{cpl1} shows how to lift any solution of the massive
6-dimen\-sio\-nal Romans' theory to a solution of 10-dimensional massive
IIA supergravity compactified on $S^3\times T^1$. This was extended to
the massless $N = 4^g$ case in ref.~\cite{nunez}, which is the one of
present interest.

The bosonic part of the relevant 10-dimensional supergravity theory is
\be 
{\mathcal L_{10}} = \tilde R -\frac{1}{2}
(\partial\phi)^2 -\frac{1}{2} e^{-\frac{1}{2}\tilde\phi} \tilde
F_{\mu\nu\rho\lambda}\tilde F^{\mu\nu\rho\lambda} \,. 
\ee
From this one can show that the right \ansatz\ to uplift our solution
to ten dimensions is given by
\bea
\label{tensix} 
d\tilde s^2_{10}& =& \frac{1}{2}e^{\frac{\sqrt{2}}{4}\phi} ds^2_6 +
\frac{1}{{\rm g}^2} e^{-\frac{3\sqrt{2}}{4}\phi} \sum_{i=1}^3
\left( \sigma^i - \frac{{\rm g}\,A^i_1}{\sqrt{2}} \right)^2
     + e^{\frac{5\sqrt{2}}{4}\phi} dZ^2\,,
\\
\tilde F_\emph{4} &=& \left( H_\emph{3} - \frac{1}{{\rm g}^2}h^1\wedge
h^2 \wedge h^3
      + \frac{1}{{\rm g}\sqrt{2}}F^i_2 \wedge h^i\right) \wedge \exd Z \,,
\\
\tilde\phi &=&\frac{1}{\sqrt{2}} \, \phi\,, 
\eea
where $h^i = \sigma^i - {1}/{\sqrt{2}}{\rm g} A^i_1$, $\sigma^i$ are
left-invariant 1-forms for $\SU(2)$. The 3-form, $H_\emph{3}$,
appearing within the expression for the 4-form is the 6-dimensional
dual~\cite{nunez},
\be
\label{f3}
H_{\mu\nu\rho} = \frac{1}{6}e^{2\sqrt{2}\phi} e\,
\epsilon_{\mu\nu\rho\lambda\beta\gamma}G^{\lambda\beta\gamma}\,,
\ee
of the 3-form field strength of the field $\exd B$ appearing in
eq.~(\ref{6daction}).

Substituting our solution with excited $B_{\mu\nu}$ field, we find for
$H_\emph{3}$
\bea
H_{x_1 x_2 x_3} &=& Q\,r^4\,e^{2\sqrt2\phi}\epsilon_{x_1 x_2
x_3}\,,
\\
    d\tilde s^2_{10}& =& \frac{1}{2\sqrt{r}} \left(-h(r)
    dt^2  + \frac{r^2\,dr^2}{h(r)} +r^2 dx^2_{1,3} + r^2
    dy^2 \right) +
\nonumber \\
&&
+\frac{1}{4 {\rm g}^2 r^{3/2}}
    \left( d\theta^2  + d\varphi^2
   + d\psi^2 + 2\cos\theta \, d\psi \, d\varphi \right)
     + \frac{1}{r^{5/2}} dZ^2\,, 
\\
 \tilde F_\emph{4} &=& Q\, e^{2\sqrt{2}\, \phi}\, r^4 \exd x_1\wedge
 \exd x_2 \wedge
 \exd x_3 \wedge \exd Z - \frac{1}{{\rm g}^2} \sin\theta \, \exd\varphi\wedge
    \exd\theta \wedge
           \exd\psi \wedge \exd Z \,,
\\
    \tilde\phi &= &-\ln r \, .
\eea

This solution can be interpreted as a D4-brane extending along the
$y$-direction, with the three angles parameterising the 3-sphere
($dx_{1,3}^2$ in the six dimensional metric), intersecting with a
D2-brane extended along the $y,Z$ directions, with the intersection
describing a string.

\subsubsection{Solutions with excited $F_\emph{2}$ or ${\mathcal F}_\emph{2}$}
These two cases can be treated together, since these fields appear in
the action~(\ref{6daction}) with the same conformal couplings: $\eta =
1/2$, $\sigma = \sqrt2 = - \lambda$, and $q = 0$ for which $n =
4$. These parameters suggest solution in Class I, for which the
$n=4$-dimensional spatial dimensions are flat, $k=0$, $\M^2 = 4$, $N =
1$ and $\rho = -1/\sqrt2$.

The solution in this case takes the form
\bea
d s_{6}^{2} &=& -h(r) d t^{2} + r^{2}d x_{0,4}^{2}
+\frac{r^{2}}{h(r)} d r^{2} \,,
\\
\phi(r) &=& -\sqrt{2} \ln r  \,,
\\
F^{\rm tr} &=&\frac{Q}{r^7}\, \epsilon^{\rm tr} \,,
\eea
where
\be 
h(r) = -\frac{2M}{r^{2}}+\frac{{\rm
g}^{2}r^{2}}{32}+\frac{Q^{2}}{8 r^6} \,. 
\ee

This geometry describes the fields due to a point source (0-brane)
in 6 dimensions, whose causal structure resembles that of an
AdS-Reissner-N\"ordstrom black hole. There can be at most two
horizons and a time-like singularity at the origin. The position
of the horizons are obtained from the positive roots of the
function $h(r)$, which in this case can be written
\be
h(r)= \frac{{\rm g}^2}{32 \,r^6} \left[r^8 - \frac{64\,M}{{\rm g}^2} r^4
                 + \frac{4\,Q^2}{{\rm g}^2}\right] =
        \frac{{\rm g}^2}{32 \,r^6} (X-X_1)(X-X_2)\,,
\ee
where $X \equiv r^4$, and the roots are
\be
    r^4_{1,2} = X_{1,2} =
    \frac{32M}{{\rm g}^2} \pm \frac{1}{2{\rm g}}
    \sqrt{\frac{(64M)^2}{{\rm g}^2}              - 16Q^2 }\,,
\ee
showing that the positions of the horizons depend on the values of $M$
and $Q$. In particular, the extremal solution is the case $16 M = {\rm
  g} Q$. An examination of the supersymmetry transformations for
Romans' gravity shows that this extremal solution is \emph{not}
supersymmetric, although the domain wall configuration for which $M =Q
=0$ reduces to the supersymmetric solution of the previous
subsubsection.

\paragraph{Uplifting procedure.}

In this subsubsection we uplift this solution to a solution of
type-IIA 10-dimensional supergravity, with nontrivial metric, dilaton
and RR 4-form, $C_{4}$. Using a slight modification of a procedure
described earlier, we get the 10-dimensional configuration
\bea
    d\tilde s^2_{10}& =& \frac{1}{2 \sqrt{r}}
    \left(-h(r)
    dt^2  +\frac{r^2\,dr^2}{h(r)} +r^2 dx^2_{0,4}  \right) +
\nonumber \\
&&
+\frac{r^{3/2}}{{\rm g}^2} \left((\sigma^1)^2  +(\sigma^2)^2 + 
\left(\sigma^3-\sfrac{{\rm g}}{\sqrt{2}}A_t^{(3)}\right)^2 \right)
    + \frac{1}{r^{5/2}} dZ^2\,,
\nn \\
    \tilde\phi &= & -\ln {r}\,, 
\\
\tilde F_\emph{4} &=& \left(- \frac{1}{{\rm g}^2} 
\sin\theta \, \exd\theta \wedge
    \exd\varphi \wedge \exd\psi
+\frac{Q}{2\sqrt{g} r^5} \exd t\wedge \exd r \wedge (\exd\psi + \cos\theta
    \, \exd\varphi) \right)
    \wedge \exd Z \,, 
\nn
\eea
where we use the conventions of the previous subsection~\ref{6exB} for
the definitions of variables. This 10D geometry describes D4 branes
along the $x_4$ directions of the 6-dimensional gauged supergravity,
plus two D2 branes extending along the $\psi$ and $\varphi$
directions.

\paragraph{T-duality.}
To relate this solution to those obtained previously, we write the
previous solution in the string frame (whose fields we label as
before with a bar). Only the metric changes, becoming
\be 
d \bar{s}_{10}^{2} = \frac{1}{2} \left[d s_{6}^2 \right]
+\frac{r^{2} }{g^{2}}\left[\sigma_{1}^{2}+\sigma_{2}^{2} +
\left(\sigma_{3}-\frac{g\,Q}{4 \sqrt{2}}\frac{1}{r^{4}} d t
\right)^{2} \right]+ r^{-2} \,d Z^{2}\,. 
\ee

This gives a solution to IIA supergravity with excited RR 4-form,
$C_\emph{4}$. We now proceed by performing a \emph{T-duality}
transformation, leading to a solution of IIB theory with
nontrivial RR 3-form, $C_\emph{3}$. The complete solution then
becomes
\bea
    d \bar{s}_{10}^{2} &=& \frac{1}{2}\, \,\left[d
    s_{6}^2 \right] +\frac{r^{2}
    }{{\rm g}^{2}}\left[\sigma_{1}^{2}+\sigma_{2}^{2} +
    \left(\sigma_{3}-\frac{{\rm g}\,Q}{4 \sqrt{2}}\frac{1}{r^{4}}
    d t \right)^{2} \right]+ r^{2} \,d Z^{2}\,, 
\\
  \bar{\phi} &= & 2 \ln {r}
\\
C_\emph{3} &=& -\frac{1}{{\rm g}^{2}} \sigma_{1} \wedge \sigma_{2}\wedge h_{3}
    -\frac{Q}{\sqrt{2} \, {\rm g}} \frac{1}{r^{5}} \, \exd t
    \wedge \exd r \wedge h_{3} \, .
\eea

We are led in this way to precisely the same 10D solution as we found
earlier --- c.f.\ formula~(\ref{7d10dsdual}) --- which we obtained by
uplifting our 7-dimensional solution to 10 dimensions.  This
establishes in detail the interrelationship between these solutions.

\subsection{Salam-Sezgin 6-dimensional gauged supergravity}

Let us now consider the chiral 6-dimensional supergravity constructed
by Salam and Sezgin in~\cite{ss}. This theory is potentially quite
attractive for applications~\cite{branesphere,6Dcc} since it has a
positive potential for the dilaton. For this supergravity we find de
Sitter-like time-dependent solutions from Class I, which are not
supersymmetric (as is expected for a time-dependent solutions).
Salam-Sezgin theory has not yet been obtained as a consistent
reduction of a higher-dimensional string theory, and so its
string-theoretic pedigree is not yet clear. What is clear is that such
a connection would be of great interest, since the solutions we find
here could be used to find interesting new string geometries with
potential cosmological applications.

The bosonic field content comprises the graviton, a 2-form potential,
$B_\emph{2}$, a dilaton and various gauge potentials, $A$, (of which
we focus on a single $\U(1)$ factor).  The bosonic lagrangian takes
the form
\be
e^{-1}{\mathcal L}_{ss} = \frac{1}{4} R -\frac{1}{4} (\partial \phi)^2
 - \frac{1}{12} e^{2\phi} G_{\mu\nu\rho}G^{\mu\nu\rho}
    - \frac{1}{4} e^{\phi} F_{\mu\nu}F^{\mu\nu} -\frac{1}{2}g^2 e^{-\phi}\,,
\ee
where $g$ is the gauge coupling and the field strengths for
$B_\emph{2}$ and $A$ are given by $F_\emph{2} = \exd A$ and $G_{\it
3} = \exd B_\emph{2} + F_\emph{2} \wedge A$. The supersymmetry
transformations for this theory can be found in~\cite{ss}. The
parameters of interest for generating solutions are $\alpha =
\beta = \frac14$, $\Lambda = g^2/2$ and $\lambda = 1$.

We do not have solutions within our \ansatze\ for which the 2-form
potential, $B_\emph{2}$, is nonzero. We do find solutions with
$F_\emph{2} \ne 0$, for which $q = 0$, $n = 4$, $\eta = \frac12$ and
$\sigma = -1 = - \lambda$. Solutions of Class I can describe this
case, for which $k=0$ and $\rho= 1$. The metric which results is
\bea
d s_{6}^{2} &=& -h(r)\, d t^{2} +\frac{r^{2}}{h(r)} d r^{2}
    + r^{2}
    d x_{0,4}^{2} \,, 
\nn\\
 \phi(r) &=& 2\, \ln r  \,,
\nonumber\\ 
F^{\rm tr} &=&\frac{Q}{r^7} \epsilon^{\rm tr} \,, 
\eea
where
\be
h(r) = -\frac{2M}{r^{2}}  - \frac{g^{2}\,r^{2}}{8}  + \frac{Q^{2}}{8\,r^6} \,.
\ee
This function has only a single zero for real positive $r$.

Since $h(r)$ is negative for large $r$, this solution is clearly
asymptotically time-dependent. Its causal structure resembles that of
a de-Sitter-S-brane spacetime (figure~\ref{noflat}$a$). Moreover, from
the supersymmetry transformations, one can show that it breaks
supersymmetry for all values of $M$ and $Q$.

\subsection{Gauged supergravity in 5 dimensions}

Romans~\cite{romansfive} has studied a gauged supergravity in 5
dimensions, corresponding to a ${\mathcal N}=4$ $\SU(2) \times \U(1)$
gauged theory. The bosonic spectrum consists of gravity, a scalar, an
$\SU(2)$ Yang-Mills potential $A^{I}$ (with field strength
$F_\emph{2}^{I}$), an abelian gauge potential $H$ with field strength
$G_\emph{2}$, and two 2-form antisymmetric potentials,
$B_\emph{2}$. Using the conventions of~\cite{tran} we consider the
reduced system without the 2-form potentials. We find for this system
two classes of point-like solutions having supersymmetric
limits. These solutions can be up-lifted, as shown in~\cite{tran}, to
solutions of ten dimensional type-II supergravity.

The action in 5 dimensions is the following~\cite{tran}:
\bea
\label{fived}
S &=& \int d^{5}x \sqrt{|g|} \Biggl[  R - \frac{1}{2}(\partial \phi)^{2}
-\frac{1}{4}e^{-\frac{2}{\sqrt{6}}\phi} F_{\mu\nu}F^{\mu\nu}
-\frac{1}{4}e^{\frac{4}{\sqrt{6}}\phi} G_{\mu\nu}G^{\mu\nu}  +\qquad
\nonumber \\&&
              \hphantom{ \int d^{5}x \sqrt{|g|} \Biggl[ }\! 
+  4 {\rm g}^{2}e^{\frac{2}{\sqrt{6}}\phi}
 -\frac{1}{4\sqrt{|g|}}\epsilon^{\mu\nu\rho\sigma\lambda}F_{\mu\nu}^I
                   F_{\rho\sigma}^I H_\lambda \Biggl]  \,,
\eea
where $G_\emph{2} = \exd H$ is the field strength for the $\U(1)$
gauge potential, $H_\mu$. We have $\alpha = 1$, $\beta = 1/2$,
$\Lambda = - 4 {\rm g}^2$ and $\lambda = -2/\sqrt6$. There are two
cases to be considered, depending on whether one of the $F^I_{\it
2}$ or $G_\emph{2}$ which is nonzero in the solution.

\subsubsection{Solutions with excited $F^{I}_\emph{2}$}\label{f5d}

In this case we have $q = 0$, $n = 3$, $\eta = 1/2$ and
$\sigma = 2/\sqrt6 = - \lambda$. These allow a 0-brane solution in
Class I, for which $k=0$, $\M^2 = 3$, $N = 1$ and $\rho = -
\sqrt2$. The resulting field configuration is given by
\bea
d s_{5}^{2} &=& -h(r) d t^{2} + r^{2}d x_{0,3}^{2}
+\frac{r^{2}}{h(r)} d r^{2} 
\label{5dfc} \\
\phi(r) &=& -\sqrt{6} \ln r 
\\
F &=&- \frac{Q}{r^4} \, \exd t\wedge \exd r \,, 
\eea
with
\be 
h(r)= -\frac{2 M}{r} +\frac{4}{9} \g^{2}r^{2}
+\frac{Q^{2}}{18\,r^4} \, ,
\ee
and where the gauge field is only nonzero for one of the gauge-group
generators.

The causal structure of this geometry is like that of an AdS-RN
black hole of positive mass, and has at most two horizons. For
negative mass there are no horizons at all and the solution has a
naked singularity at $r = 0$. For $M > 0$ the extremal limit
corresponds to the case where the two roots of $h(r)$,
\be
r^3_{1,2} =\frac{9\,M}{4\,\g^2} \pm
\frac{1}{2}\sqrt{\left(\frac{9\,M}{2\,\g^2}\right)^2 -
  \frac{Q^2}{2\,\g^2}}\, ,
\ee
coincide, corresponding to when ${9\,M}/{2\,\g} = {Q^2}/{\sqrt{2}}$,
and for which the solution is static everywhere but is \emph{not}
supersymmetric.

A supersymmetric configuration is obtained in the totally static,
uncharged case, corresponding to the choices $M = Q^{2} = 0$. This
represents a domain-wall-like object and preserves one of the
supersymmetries of the action.

\paragraph{Uplifting procedure to 7 dimensions.}
 In this subsection we show how the lift of this 5D solution to 7
 dimensions on 2-torus gives \emph{the same} solution as we found in
 subsubsection~(\ref{sim7sol}). This reinforces the conclusion that
 the solutions we find are related to one another by dimensional
 reduction/oxidation, or through dualities.

To this end we denote 7-dimensional quantities with a tilde, and
rescale the coordinates in the following way
\be 
\tilde{t}=\sqrt{\frac{25}{9}}  t \, , \qquad
\tilde{r}=r^{{3}/{5}} \, , \qquad 
d \tilde{x}_{k,3} =d x_{k,3} \,. 
\ee
The uplifting procedure acting on the solutions then gives~\cite{tran}
\bea
d \tilde{s}_{7}^{2} &=& e^{\frac{4}{5\sqrt{6}} \phi} d s_{5}^{2}
+e^{-\frac{6}{5\sqrt{6}} \phi}(d\, Y^{2}+d\, Z^{2})  
\nn\\
\tilde{\phi} &=& \sqrt{\frac{6}{10}} \, \phi 
\nn\\
\tilde{G}_{4} &=& F_{2} \wedge \exd Y \wedge \exd Z   
\nn\\
\tilde{F}_{2}^{I} &=& F_{2}^{I} \,. 
\eea
Applied to the previous formulae, with the redefinition
\bea 
\tilde{h}(\tilde{r}) &=& \frac{9}{25} 
\tilde{r}^{-\frac{4}{3}} h(\tilde{r}) 
\nn \\
&=&-\frac{2\tilde{M}}{\tilde{r}^{3}}+\frac{4 \g^{2}
\tilde{r}^{2}}{25} +\frac{Q^{2}}{50 \, \tilde{r}^{8}} 
\eea
one ends with the following 7-dimensional metric
\be
 d \tilde{s}_{7}^{2}=-\tilde{h}(\tilde{r}) d \tilde{t}^{2} +
\tilde{r}^{2}d \tilde{x}_{0,5}^{2}
+\frac{\tilde{r}^{2}}{\tilde{h}(\tilde{r})} d \tilde{r}^{2} \, .
\ee
This is exactly the same form found in subsection~(\ref{7dtv}). (It is
straightforward to check that the other fields also transform to the
fields found in the 7-dimensional case.)

\paragraph{Uplifting procedure to 10 dimensions.}
 As discussed in~\cite{tran}, any solution to Romans' 5-dimensional
 gauged supergravity~(\ref{fived}) can be lifted to a solution of
 10-dimensional type-II string theory, with the solutions so obtained
 corresponding to a 5-brane wrapped in a non-trivial way.

Let us therefore consider type-IIB supergravity in 10 dimensions, for
which the relevant bosonic part of the truncated action is given by
\be
\label{type-IIB} 
{\mathcal L_{10}} = \tilde R -
\frac{1}{2}(\partial\tilde \phi)^2 -
\frac{1}{2}e^{-\tilde\phi}\tilde F_{\mu\nu\rho}
\tilde F^{\mu\nu\rho}\,. 
\ee
The complete reduction \ansatz\ for the 10-dimensional solution
compactified on $T^1 \times S^3\times T^1$ is given by~\cite{tran}
\bea
\label{tend}
 d\tilde s^2_{10} &=& e^{\frac{13}{5\sqrt{6}}\phi}ds^2_5 +
e^{\frac{3}{5\sqrt{6}}\phi} (dY^2 + dZ^2) + \frac{1}{4\g^2}
e^{\frac{-3}{\sqrt{6}}\phi}\sum_{i=1}^3(\sigma^i - \g A_1^i)^2 \,, 
\\
 \tilde F_\emph{3} &=& e^{\frac{4}{\sqrt{6}}\phi}*G_\emph{2} -
\frac{1}{24\,\g^2}\epsilon_{ijk}\, h^i\wedge h^j \wedge h^k +
\frac{1}{4}\sum_{i=1}^3 F_\emph{2}^i\wedge h^i \,, 
\\
 \tilde \phi &=& \sqrt{6} \,\phi  \quad {\rm and} 
\quad h^i =  \sigma^i - \g\,A_\emph{1}^i\,.
\eea
Since we choose $A_\emph{1}^i= A_t^3 = A_t$ and $G_{\it 2}=0$ for the
present purposes, the 10-dimension solution reduces to
\be
 h^1 = \sigma^1 \,, \qquad  h^2 =  \sigma^2 \,,
\qquad  h^3 = \sigma^3 - {\rm g} A_t \,, 
\ee
and so
\bea 
\phi &=& - \ln (r)
\\
d\tilde s^2_{10} &=& \frac{1}{r^{13/5}}
\left[ -h(r) d t^{2} + r^{2}d x_{0,3}^{2}
+\frac{r^{2}}{h(r)} d r^{2}\right] + \frac{1}{r^{3/5}} (dY^2 + dZ^2) +
\nonumber\\ 
&& + \frac{r^3}{4\g^2} \left[\sigma_1^2
+ \sigma_2^2 + \left(\sigma_3^2 - \sfrac{A_0}{r^3}dt\right)^2\right] , 
\nonumber\\ 
\tilde F_\emph{3} &=& -\frac{1}{4\,\g^2}\,\sin\theta \, \exd\theta
\wedge \exd\varphi \wedge \exd\psi
 + \frac{Q}{4r^4} \, \exd t\wedge \exd r \wedge
( \exd\psi + \cos\theta \, \exd\varphi)\,. 
\eea
In these expressions $\sigma_i$ are defined as before (see
eq.~\ref{sigmas}) This 10-dimensional solution to type-IIB
supergravity describes an NS-5 brane, extending along the $x_1, x_2,
x_3, Y, Z$ directions, plus two F1 strings extending along the $\psi,
\varphi$ directions.

\subsubsection{Solutions with excited $G_\emph{3}$}\label{s5d}

In this case, we have $q = 0$, $n = 3$, $\eta = 1/2$ and $\sigma =
-4/\sqrt6 = 2 \lambda$, and so we can obtain solutions of Class
III. For these $k = c = 1$, $\M^2 = 3$, $N = 1$ and $\rho =
-\sqrt{2}$. The metric becomes
\bea
d s_{5}^{2} &=& -h(r) d t^{2} + r^{2}d x_{1,3}^{2}
+\frac{r^{2}}{h(r)} d r^{2} 
\label{5dsc} \\
\phi(r) &=& -\sqrt{6} \ln r 
\\
G_{\rm tr} & = & -Q\, r^2 dt\wedge dr
\eea
with
\be
h(r)= -\frac{2 M}{r} +\frac{r^2}{9} (4\,\g^2 + Q^2)\,.
\ee
This solution has a single event horizon at $r_h^3= {18M}/
{(4\g^2+Q^2)}$. It has the same causal structure as an
AdS-Schwarzschild black hole.

This solution turns out to be supersymmetric if we choose $M$ and $Q$
equal to zero. This is the same supersymmetric solution found in
subsection~\ref{f5d} in these limits.

\paragraph{Uplifting procedure to 10 dimensions.}
This solution may also be lifted using \linebreak
eqs.~(\ref{type-IIB}),~(\ref{tend}) into a full solution to type-IIB
supergravity on $T^1 \times S^3\times T^1$. In the present instance,
$F_2^I = 0$ and so $h^i=\sigma^i$ which implies
\bea 
\phi &=&-\ln(r) 
\nonumber\\ 
d\tilde s^2_{10} &=& \frac{1}{r^{13/5}}\left[-h(r) d t^{2}
+ r^{2}d x_{1,3}^{2}  +\frac{r^{2}}{h(r)} d r^{2}\right] +
\frac{1}{r^{3/5}} (dY^2 + dZ^2) +
\nn \\
&& + \frac{r^3}{4\g^2}
[d\theta^2  +d\varphi^2
 +d\psi^2 + 2 \cos\theta \, d\psi \, d\varphi]
\nn\\
\tilde F_\emph{3} &=& \frac{Q}{r^4} \sin^2\alpha \sin\,\beta \,
\exd\alpha\wedge \exd\beta\wedge \exd\gamma
 -\frac{1}{24\,\g^2}\,\sin\theta \sin\psi \, \exd\theta \wedge \exd\varphi
 \wedge  \exd\psi \,.
\eea
Here we use the metric for the 3-sphere (inside the 5 dimensional
gauged supergravity) to~be
\be 
dx_{1,3}^2= d\alpha^2 + \sin^2\alpha \, d\beta^2 +
\sin^2\alpha \sin^2 \beta \, d\gamma^2 
\ee
and we use the same conventions of subsubsection~(\ref{f5d}) in the
definition of the variables. We obtain in this case a solution
representing two NS-5 branes that intersect in the $Y,Z$ directions.

\subsection{Non-compact gauged  supergravity}

The absence of de Sitter-like solutions to gauged $N=8$ 4D
supergravity using compact gaugings has led to the study of
non-compact and non-semi-simple gaugings, using groups like
$\CSO(p,q,r)$, such as developed in~\cite{hull}. These in some cases
can have de Sitter solutions. When regarded as dimensional reductions
from a higher-dimensional theory, these theories correspond to
reductions on hyperboloids, which give $\SO(p,q)$ gaugings. Recent
effort has been devoted to the study of possible cosmological
applications of non-compact gauged supergravities~\cite{linde}. In
this subsubsection we present an example of non-compact gauged
supergravity in four dimensions which furnishes alternative examples
of the cosmological solutions we have been considering, rather than de
Sitter space.

We consider the 4-dimensional model studied in~\cite{ahn} (see
also~\cite{linde}) with $\SO(a,b)$ gauging, where $a$ and $b$ are
natural numbers satisfying $a+b=8$. Let us consider a consistent
truncation of the complete supergravity action describing gravity
coupled to a single real scalar field:
\be
\label{ncogauge} 
S=\int d^{4}x \sqrt{|g|}
\left(R-\frac{1}{2}(\partial \phi)^{2} -V_{a,b}(\phi) \right) 
\ee
where the potential $V_{a,b}$ is given by
\be
\label{poncogauge}
V_{a,b}=\frac{{\rm g}^2}{16}a \left(b-3 a \right)
e^{\sqrt{\frac{b}{a}} \phi} \, .
\ee
This action falls into our general class, with $\alpha = 1$, $\beta =
1/2$, $\Lambda = {1}/{16} \, {\rm g}^2 a(b - 3a)$ and $\lambda
= - \sqrt{b/a}$. The absence of a antisymmetric form requires the
choice $Q = 0$.

This action admits solutions for each class, although it is possible
to see that these all coincide. For this reason, we limit our
discussion to Class I, for which $k = 0$ and $n + q = 2$. If we also
choose $c = 1$ (for which we have no choice if $q \ne 0$) then $\M^2 =
2$ and $N = 1$, leading to $\rho=-\sqrt{{2b}/{a}}$.  The coefficients
of the metric are given by
\be
\label{hncogauge}
h(r)=-2M r^{\frac{b}{a}-1}+\frac{{\rm g}^{2}}{32}a^{2} r^{2} 
\ee
and
\be 
g(r)=r^{-\frac{2b}{a}}h(r) 
\ee
while the scalar field is
\be 
\phi(r)=-2\, \sqrt{\frac{b}{a}} \, \ln(r) \, .
\ee

It is interesting to see that this solution corresponds to the
particular example we studied in subsection~(\ref{specprop}), and all
the properties discussed in that case also apply here. In particular,
notice that we have asymptotically time-dependent solutions for $b >
3a$, when the parameter $M$ is positive, even though the scalar
potential is negative (AdS-like). The resulting geometry is
asymptotically flat, with a Penrose diagram corresponding to the
S-brane of figure~\ref{sbrpen}.

\section{The general case}\label{part4}

We now develop our earlier analysis more generally, with an eye to its
application to theories having more general scalar potentials.

\subsection{Systematic derivation and generalizations}\label{appA}

This subsection is devoted to presenting the general procedure we use
to obtain the field configurations of subsection~\ref{123456}, and to
several examples of its use in finding new solutions for different
potentials. We believe that the procedure we shall follow --- which is
a generalization of the one presented in~\cite{bowcock,langlois} ---
is particularly suitable for exhibiting the relations between the
choice of the scalar potential and the characteristics of the
solution.

\subsubsection{Derivation: the general method}

We start by considering the general action of~(\ref{generalaction}),
and the following \ansatz\ for a metric that depends on two
independent coordinates:
\be 
ds^2 = 2\, e^{2
B(\tilde t,\tilde r)}(- d\tilde t^2 +  d\tilde r^2) + e^{2
A(\tilde t,\tilde r)}
           dx_{k,n}^2  +  e^{2 C(\tilde t,\tilde r)}dy_{q}^2\,.
\ee
\looseness=-1 Here, as before, $ dx_{k,n}^2$ describes the metric of an
$n$-dimensional maximally-symmetric, constant-curvature space, with
parameter $k=\pm1,0$. The $q$ flat dimensions identify the spatial
directions parallel to the $(q+1)$-dimensional extended object. We
further assume this object to carry electric charge for the
$(q+1)$-dimensional antisymmetric gauge potential.

It is convenient to use double null coordinates, $u = t - r$ and $v =
t + r$ throughout what follows, since for these the field equations
take a simpler form. With these choices the metric takes the form
\be 
ds^2 =  -2\,e^{2 B(u,v)} \, dudv +  e^{2 A(u,v)} \, dx_{k,n}^2
           +  e^{2 C(u,v)} \, dy_{q}^2      \,.
\ee
Assuming that the $(q+2)$-form field strength depends
only on the coordinates $(u,v)$, its field equation can be readily
integrated to give
\be
\label{F} F^{uvy_1\dots y_q}= Q\, e^{\sigma\phi - 2B-nA - qC}
         \,    \epsilon^{uvy_i\dots y_q}
\ee
where $Q$ is a constant of integration, which we interpret as the
electric charge, and $\epsilon^{\ldots }$ is the usual antisymmetric
tensor density whose elements are $\epsilon^{uvy_1\dots y_q} = 1$.

The Einstein equations corresponding to the $(u,v)$, $(u,u)$,
$(v,v)$, $(i,j)$ and $(q,p)$ directions,
together with the equation for the dilaton, become
\bea
\label{uu} 
&& 
2n A_{u}B_u +2q C_{u}B_u  = nA_{uu} + nA_u^2 +
qC_{uu} + qC_u^2 + \frac{\beta}{\alpha} \phi_u^2\,,
\\
&& 2n A_{v}B_v +2q C_{v}B_v  = nA_{vv} + nA_v^2 +
qC_{vv} + qC_v^2 +
\sfrac{\beta}{\alpha} \phi_v^2\, 
\label{vv}\\
&& 
n A_{uv} + n^2\,A_{u}A_{v} + q C_{uv} +q^2 C_u C_v + nq (A_uC_v + A_vC_u)=
\nonumber\\
&& \qquad\qquad    
= -\frac{n(n-1)\,k}{2}  e^{2(B-A)} 
+ \frac{\,\eta\, Q^2}{2\alpha} e^{\sigma\phi + 2(B-nA)}
+ \frac{V(\phi)}{2\alpha} e^{2B}\,,
\label{uv}\\ 
&&  
2B_{uv} {+} 2(n{-}1)A_{uv} {+} n(n{-}1)A_uA_v {+}2qC_{uv}
 {+}(nq{-}q)[A_vC_u {+} A_uC_v] {+}q^2 C_uC_v + q C_uC_v =\quad
\nonumber \\
&& \qquad\qquad 
= - \frac{\beta}{\alpha}\phi_u\phi_v
- \frac{(n-1)(n-2)}{2}k e^{2(B-A)}  -
\frac{\eta\,Q^2}{2\alpha}e^{\sigma\phi + 2(B-nA)}
+ \frac{V}{2\alpha}e^{2B}\,, 
\label{ii}\\ 
&& \label{yy}
 2B_{uv}{+} 2nA_{uv} {+} n(n{+}1)A_uA_v {+}2(q{-}1)C_{uv} 
{+}n(q{-}1)[A_vC_u {+} A_uC_v] {+}q(q{-}1)C_uC_v  =
\nonumber \\
&& \qquad\qquad
= - \frac{\beta}{\alpha}\phi_u\phi_v
- \frac{n(n-1)}{2}k e^{2(B-A)}  +
\frac{\eta\,Q^2}{2\alpha}e^{\sigma\phi + 2(B-nA)} + \frac{V}{2\alpha}e^{2B}\,,
\\ \nonumber \\
\label{scaleq}
&& 2\phi_{uv}+n A_{v}\phi_{u} + n A_{u}\phi_{v}+q C_{v}\phi_{u}+q
C_{u}\phi_{v} = -\frac{\sigma \eta Q^{2}}{2 \beta}
e^{\sigma\phi+2(B-nA)} - \frac{V'(\phi)}{2 \beta} e^{2 B}\,.
\eea
where we use~(\ref{F}) to rewrite eqs.~(\ref{uv}--\ref{scaleq}).  As
before $\alpha$, $\beta$, $\eta$ and $\sigma$ denote the constants
which appear in the action \pref{generalaction}. In these equations
the sub-indices denote differentiation with respect to the
corresponding variable. As usual, the Bianchi identity ensures that
one combination of the previous equations is redundant. Because of the
presence of the dilaton field, Birkhoff's theorem does not preclude
solutions depending on both coordinates $u,v$. At a later point we
specialize in solutions that depend only on a single coordinate.

We call the generalization of our previous \ansatze\ to the present
case ``proportionality'' \ansatze, and they have the form
\be
\label{ansone}
C(u,v) = c A(u,v)\,,
\ee
where  $c$ is  a constant, and we ask the dilaton to be an
arbitrary  function of $A$, only:
\be
\label{anstwo} \phi(u,v) =
\mathcal M \,S(A(u,v)) 
\ee
where as before $\mathcal M^2= n + cq$.

\paragraph{Finding the solutions.}
We first concentrate on the Einstein equations, for which use of the
\ansatze~(\ref{ansone}) and~(\ref{anstwo}) in eqs.~(\ref{uu})
and~(\ref{vv}) imply the function $A$ must take the form
\be
A(u,v) = f( \mathcal U(u) + \mathcal V(v))\,,
\ee
where $f$, $\mathcal U$ and $\mathcal V$ are arbitrary functions of a
single variable. With this information we can integrate~(\ref{uu})
and~(\ref{vv}) to find the functional form for $B(u,v)$ is
\be
\label{B}
2\, B(u,v) = N f + L(f) +
\ln|f'(\mathcal U(u)+ \mathcal V(v))| + \ln|\mathcal U'(u) \mathcal
V'(v)|  + \ln \xi
\ee
where  $\xi$ is a constant of integration,
\be 
L(A)= \frac{\beta}{\alpha}\int{(S')^2 \,df}\,, 
\ee
and, as before, $N= (n+ c^2 q)/{\mathcal M^2}$.

Using~(\ref{B}) we can now write~(\ref{uv}) as the following
differential equation for $f$,
\be
\label{fprima}
\mathcal U'\mathcal V'\left[ f'' + \M^2\, f'^2 +  f'  F_1(f)\right] =0\,,
\ee
where
\be
 F_1(f)\equiv  \xi  \, e^{N\,f + L(f)}
\left(\frac{n(n-1)k}{2 \mathcal M^2} \, e^{-2 f}
 - \frac{\eta \, Q^2}{2\alpha \mathcal M^2} \, e^{\sigma
\mathcal M S(f)- 2nf}  
  - \frac{V(S(f))}{2\alpha \mathcal M^2} \right).
\ee
In all of these expressions a prime denotes differentiation with
respect to the function's only argument. Assuming $\mathcal U'$ and
$\mathcal V'$ are both nonzero, this relation gives a second order
differential equation for $f$, whose first integral gives
\be
\label{fprimasol}
  f'\, e^{\mathcal M^2 f} = - \int e^{\mathcal M^2 f} F_1(f) \, df
+  2M \, , 
\ee
where $2M$ is an integration constant.

Let us consider next the $(ii)$ and $(yy)$ components ---
eqs.~(\ref{ii}) and~(\ref{yy}) --- of the Einstein equations.  (Recall
that for point-like objects ($q=0$) and so eq.~(\ref{yy}) need
\emph{not} be imposed. In this case eq.~\pref{ii} also need not be
separately imposed inasmuch as the Bianchi identity makes it not
independent of those we consider explicitly. However, for $q \neq 0$
(brane-like objects) only one of these two equations is dependent on
the others, leaving the other to be solved explicitly.)

It is convenient to consider the independent equation to be the
difference of eqs.~(\ref{ii}) and~(\ref{yy}) (with the Bianchi
identity making the sum redundant). With our \ansatze, one finds
the following differential equation for $f$:
\be
\label{fprimadue}
 \mathcal U'\mathcal V'\left[ (c-1)\left(
 f'' + \M^2\, f'^2\right) +  f'  G_1(f)\right] =0\,,
\ee
where
\be
G_1(f)\equiv  \xi  e^{N\,f + L(f)}
\left( (n-1)\,k e^{-2f} - \frac{\eta Q^2}{\alpha}e^{\sigma \M S(f)-2nf}
\right)\,.
\ee
Comparing eq.~(\ref{fprimadue}) with~(\ref{fprima}) leads to the
following new condition
\bea
\label{constrayygen}
    && q\cdot\left[ (n-1)\,k- \frac{\eta
    Q^2}{\alpha}e^{\sigma \M S(f)-2(n-1)f} \right] =
    \nonumber \\
    && \qquad\qquad q\cdot (c-1)\left[ \frac{n(n-1)\,k}{\M^2 } - \frac{\eta
    \, Q^2}{\alpha \M^{2}} \,
    e^{\sigma \M S( f) -2(n-1)f } -\frac{V}{\alpha \M^{2}} \, e^{2 f}
    \right] \eea
where the explicit factors of $q$ show that this constraint only holds
when $q \neq 0$.

Next we substitute eq.~(\ref{B}) into the scalar
equation,~(\ref{scaleq}), to obtain the following relationship between
$S'(=dS/df)$ and $f'(=df/d(\mathcal U + \mathcal V))$
\be
\label{sprima}
 \mathcal U'\mathcal V'\left[ S'\,f'' + \M^2\, S'\, f'^2 + S''\,f'^2
 + f' F_2(f)\right] =0\,,
\ee
where
\be 
F_2(f) \equiv  \frac{\xi}{2}\, e^{N\,f + L(f)} \left(
\frac{\sigma\,\eta\,Q^2}{2\beta \mathcal
M} \, e^{\sigma \mathcal M S(f) - 2nf} 
+ \frac{V'(S)}{2\beta \mathcal M^2} \right) .
\ee
Direct differentiation gives $\frac{d}{dx} (S') = f' S''$, where $f =
f(x)$, which after using eq.~(\ref{sprima}) implies:
\be 
\frac{d}{dx} (S')  = S' F_1 -F_2  \, . 
\ee
Combining this equation with~(\ref{fprima}) to eliminate $f'$ allows
us to obtain the following equation for $S(f)$ purely in terms of
$F_1$ and $F_2$:
\be
\label{sdev} 
S' F_1-F_2 = S''e^{-\mathcal M^2 f}
\left(-\int {\rm e}^{\mathcal M^2 f}F_1 df + 2M \right) 
\ee

Now comes the main point. Given any particular explicit scalar
potential, $V(\phi)$, eq.~(\ref{sdev}) may be regarded as a
differential equation for $S$, which may be (in principle) explicitly
integrated. Given the solution we may then use~(\ref{fprima}) to find
$f'$. There are two cases, depending on whether or not $S''$ vanishes.

\begin{enumerate}
\item
If $S' = \rho$ is a constant, then the equation for $f$ becomes $F_2 =
\rho F_1$.

\item
For $S''\neq 0$ it is possible to explicitly write the non-linear
differential equation satisfied by $S$, which follows
from~(\ref{sdev}). The result is
\be
\label{nonlinear} 
S'''\left(F_2-S'F_1\right) + 2S''^2 F_1 -
S''\left(F_2'+\mathcal M^2 F_2\right) + S'' S'\left(F_1' +
\mathcal M^2 F_1 \right)= 0  \, .
\ee
Notice that the dependence on $L=\frac{\beta}{\alpha}\int S'^2 df$
disappears in this equation since it only enters as an overall
exponential factor in both $F_1$ and $F_2$, so eq.~\pref{nonlinear} is
a genuine differential equation rather than an integro-differential
equation.
\end{enumerate}

Knowing the potential, $V(S)$, we can explicitly find $F_1$ and $F_2$
and our problem reduces to the solution of one single differential
equation, eq.~(\ref{nonlinear}). In principle this can always be done,
even if only numerically. Alternatively, since eq.~\pref{nonlinear} is
complicated, in the examples to follow we will take the inverse route
where we choose a simple function $S(f)$ and then determine which
scalar potential would be required to make this function a solution.

Once we know $f$ (using the other Einstein equations) we can in
principle then find the metric everywhere, using
\be
ds^2 = -2\xi\,|\mathcal U'\mathcal V'|f'e^{Nf+ L(f)}dudv
      + e^{2 f} dx_{k,n}^2  + e^{2 c f} dy_{q}^2\,.
\ee
where we now take $\xi=2$ for convenience.

Although the metric appears to depend independently on two
coordinates, it really only depends on a single coordinate given
our \ansatze. To see this it is useful to write the metric in the
following coordinates
\be
r = e^f  \,, \qquad t= \mathcal U - \mathcal V\,,
\ee
since, with these coordinates, the general solution takes a form
depending only on the single coordinate $r$:
\be
ds^2 =  -h(r)\, dt^2 +  \frac{dr^2}{g(r)} + r^2 dx_{k,n}^2
             + r^{2c} dy_{q}^2\,,
\ee
with
\be \label{ache}
h(r) =- f'\,  r^N e^{L(\ln r)}
\ee
and
\be
\label{ge1} g(r)= h(r)\, r^{-2(N  -1)} e^{-2L(\ln r)}\,. 
\ee
We arrive in this way exactly the same form for the solution as found
in subsection~\ref{123456} by directly integrating the equations of
motion. The explicit form for $f'$ appearing here may be read off from
eq.~(\ref{fprimasol}). The procedure from here on proceeds as before:
by comparing the above with the scalar equation we obtain two
different expressions for $h$, whose consistency constrains the
model's various parameters.

To summarize, we can obtain the form of $S(\ln r)$ from~(\ref{sdev})
and once we have solved for $S$, given an explicit scalar potential,
we can easily find $L(\ln r )$ and from this determine the full
metric. In practice however, the solution of this equation is very
difficult.

The real advantage to the derivation of this section is the ability to
set up and solve the inverse problem, wherein we choose a particular
ansatz for $S(\ln r )$ and find the scalar potential which gives rise
to such a solution. We illustrate this technique with two non-trivial
examples in the next two sections.

\subsection{Solutions with sums of Liouville terms}\label{maslambda}

Non-trivial solutions can be found by taking the simplest possible
choice: $S'' = 0$, or
\be
\label{anssl} 
S(f)=\rho f 
\ee
for some constant $\rho$. This choice leads to the following for
$\phi$
\be
\label{ansmaslam}
\phi(f)= \mathcal M S(f)= \rho \mathcal M f\,.
\ee

With this simple form, the first terms of eqs.~(\ref{fprima}),
and~(\ref{sprima}) become equal and so one can already find the
general form of the potential that satisfies the solutions by just
looking for the solutions of
\be 
F_2 = \rho F_1 \,.
\ee
This leads to the following differential equation for the
potential
\be
\label{V1} 
V'(f) - \Gamma V(f) = -\Upsilon\, e^{-2f} + \Theta\, e^{\chi f} 
\ee
where
\be 
\Gamma= \frac{2\beta \,\rho^2}{\alpha}\,, 
\qquad 
\Upsilon = 2\beta\, \rho^2\,  n(n-1)k\,, 
\qquad
\chi = \sigma \mathcal M \rho -2n\,,
\ee
and
\be
\Theta = Q^2\,\eta\, \rho \mathcal M \left[ \frac{2\beta
\rho}{\alpha \mathcal M} + \sigma \right]\,,
\ee
eq.~(\ref{V1}) is solved by writing the potential as $V(f) =
G(f)e^{\Gamma f}$, where $G(f)$ is an arbitrary function of $f$.
Plugging this into~(\ref{V1}) one finds
\be G(f) =
\sfrac{\Upsilon}{\Gamma + 2} \, e^{-(\Gamma + 2) f}
 + \sfrac{\Theta}{\chi - \Gamma}\,e^{(\chi - \Gamma) f} + V_0\, \,,
\ee
from which we conclude that the potential is
\be
\label{V2}
V(\phi) = \sfrac{\Upsilon}{\Gamma + 2}\,e^{-(2/\rho \mathcal M)\phi}
+ \sfrac{\Theta}{\chi - \Gamma}\,e^{(\chi/\rho \mathcal M)\phi }
+ V_0\, e^{(\Gamma/\rho \mathcal M)\phi}\,,
\ee
where $V_0$ is an integration constant. This shows that the most
general form for $V$ consistent with our \ansatz~(\ref{ansmaslam}), is
the sum of \emph{no more than three} exponentials. Of course, in order
to get the full solutions we also must impose the
constraint~(\ref{constrayygen}) if $q \ne 0$.

To determine the explicit form for the solutions in this case we
follow a straightforward generalization of what was done above for the
Liouville potential. This involves comparing the two different forms
for $h$ which are obtained from the Einstein and dilaton equations.
At this point, the procedure is exactly the same as in the case of
single Liouville potential, with the substitution of the single
exponential $\Lambda \, e^{-\lambda \phi}$ with the sum $\Lambda_i \,
\ e^{-\lambda_{i} \phi}$.  For $q\ne 0$ we shall find in this way that
there are two classes of solutions, having at most two exponentials in
the potential (as can be seen by looking at eq.~(\ref{constrayygen})),
whereas for $q=0$ a third class is allowed having three terms in the
dilaton potential.

We are led in this way to the following 5 classes of solutions, the
first two of which can arise with non vanishing $q$.

\paragraph{Class I$_i$.}
This class contains solutions with $k \ne 0$. In order to have two
terms in the potential for the dilaton, we must require the terms
in $h(r)$ proportional to $k$ and $\Lambda_1$ to have the same
power of $r$, and separately require the same of those terms
proportional to $Q^2$ and $\Lambda_2$. This leads to the following
relations amongst the parameters:
\be 
\label{class1iq}
\left\{
\begin{array}{l}
\displaystyle
\lambda_1 \rho \M = 2 \,, \\[3pt]
\displaystyle
\rho\M = \frac{2n}{(\sigma + \lambda_2)} \,, \\ [8pt]
\displaystyle
\Lambda_1 \left[ \frac{1}{\alpha \M} -
\frac{\lambda_1}{2\beta\rho}\right] = \frac{n\,(n-1)\,k}{\M}\,, \\[8pt]
\displaystyle
\eta\,Q^2 \left[ \frac{1}{\alpha \M} +\frac{\sigma}{2\beta\rho}\right] =
   \Lambda_2 \left[ \frac{\lambda_2}{2\beta\rho} -\frac{1}{\alpha \M}
                      \right]  \, \\[8pt]
\displaystyle
q\cdot\left[ \alpha\,(n-1)\,k\,(n\,[c-1]-\M^2)=
                       \Lambda_{1} \,(c-1)\right]\,, \\ [3pt]
\displaystyle
q\cdot\left[ -\eta\, Q^2(c-1-\M^2)= (c-1) \Lambda_2 \right] \,.
\end{array} \right.
\ee
The last four of these relations gives the expression for $Q^2$
and $\Lambda_i$ in terms of the other parameters. In this case the
function $h$ becomes
\bea
\label{class1iqh}
h(r)&=& -2 M r^{N-\M^{2}+ \beta\rho^2/\alpha } -
\frac{\lambda_1\Lambda_1 \, r^{2N + 2\beta\rho^2/\alpha-2}}
{2\beta\rho \M [\M^2-2+ N +\beta\rho^2/\alpha]} +
\nonumber \\ 
&& + \frac{[\sigma \, \eta\,Q^2 -\lambda_2 \Lambda_2]\,
     r^{2N + 2\beta\rho^2/\alpha-\lambda_2\rho \M}}
   {2\beta\rho\M\,[\M^2  +N +\beta\rho^2/\alpha -\lambda_2\rho\M]} \,.
\eea
The solutions for $q=0$ in this class are easily obtained as
special cases of the constraints given above.

\paragraph{Class II$_i$.}
This class also nontrivial solutions for $k\ne 0$. In this case we
demand the terms in $h(r)$ proportional to $k$, $\Lambda_1$ and
$Q^2$ all share the same power of $r$, and leave the $\Lambda_2$
term by itself. In order to satisfy eq.~(\ref{constrayygen}) we
then require
\be \label{class2iq}
\left\{
\begin{array}{l}
\lambda_1 \rho \M = 2 \,, \\[3pt]\displaystyle
\rho\M = \frac{2n}{(\sigma + \lambda_1)} \,, \\ [8pt]
2 \beta \rho = \lambda_2 \alpha \M \,, \\ [3pt]
\eta\,Q^2 \left[ \lambda_2 + \sigma \right] +
   \Lambda_1 \left[ \lambda_2- \lambda_1 \right] =
                        \alpha\lambda_2 \,n\,(n-1)\,k \,, \\[3pt]
 c=1 \,, \\ [3pt]
 q \cdot [ \Lambda_1 \left[ \lambda_2 - \lambda_1 \right] =
           \alpha\,(n-1)\,k [\lambda_2 (n-1) -\sigma]]   \,.
\end{array} \right.
\ee
In this case the function $h$ becomes
\be\label{class2iqh}
h(r)=-2 M r^{1-\M^{2}+ \beta\rho^2/\alpha } +
\frac{ \Lambda_2 \, r^{2}}
{\alpha\M^2 [\M^2 + 1 -\beta\rho^2/\alpha]} 
+ \frac{[\sigma \, \eta\,Q^2 -\lambda_1 \Lambda_1]\,
     r^{2\beta\rho^2/\alpha}}
   {2\rho\beta\M\,[\M^2 -1 +\beta\rho^2/\alpha]} \,.
\ee
The solutions for $q=0$ are again easily obtained as special cases of
the constraints above. The next classes of solution are only possible
for $q=0$.

\paragraph{Class III$_i$.}
This class requires $k= q = 0$ and allows two terms in the potential
provided the parameters satisfy the relations
\be
\left\{
\begin{array}{l}
(\sigma + \lambda_1)\sqrt{n} \rho  = 2n \,, \\ [3pt]    
\alpha \sqrt{n}\lambda_2 =  2 \beta \rho   \,, \\ [3pt]
\eta\,Q^2 [\sigma + \lambda_2 ] = \Lambda_1 [\lambda_1 - \lambda_2]
\end{array} \right.
\ee
These conditions imply the constraint $\lambda_2 = \frac{4\beta
n}{\alpha n (\sigma + \lambda_1)}$. The function $h$ in the
metric in this case becomes
\be
\label{class1v}
h(r) = -2 M r^{1-n+ \beta\rho^2/\alpha } -
\frac{\Lambda_2\, r^{2}}{\alpha \,n\,[n +1-\beta\rho^2/\alpha]}
+\frac{[\sigma\,\eta\,Q^2 - \lambda_1
\Lambda_1] \,\, r^{2(\beta\rho^{2}/\alpha+1)}}{2\beta\rho\sqrt{n}\,[n + 1
  +\beta\rho^2/\alpha -\lambda_1\rho\sqrt{n}]\,r^{\lambda_1\rho\sqrt{n}}}\,.
\ee

\paragraph{Class IV$_i$.}
This class contains solutions valid only for $q = 0$ but for any
curvature $k$, and allows two terms in the potential. (They
correspond to combining together the term in $\Upsilon$ and that
with $\Theta$.) The parameters must satisfy the following
constraints:
\be \label{classiiv}
\left\{
\begin{array}{l}
\lambda_1 \rho \sqrt{n}= 2 \,, \\[3pt]
2\beta\rho = - \alpha \sigma\,\sqrt{n} \,, \\ [3pt]
2\beta \rho= \alpha \lambda_{2} \sqrt{n} \\ [3pt]
\displaystyle 
\Lambda_1 = \frac{\alpha \sigma \,n(n-1)\,k}{\sigma + \lambda_1}\,.
\end{array} \right.
\ee
This implies that $\lambda_2 = -\sigma$.  In this case the function
$h$ becomes
\bea
\label{class2v}
h(r)&=& -2 M r^{1-n+ \beta\rho^2/\alpha } -
\frac{\lambda_1\Lambda_1 \, r^{ 2\beta\rho^2/\alpha}}
{2\beta\rho \sqrt{n} [n-1 +\beta\rho^2/\alpha]} -
\nonumber \\
&& - \frac{\Lambda_2\, r^{2}}{\alpha \,n\,[n  +1
-\beta\rho^2/\alpha]} + \frac{\eta\,Q^2}
 {\alpha n [n - 1 + \beta\rho^2/\alpha]\,r^{2(n-1)}}\,.
\eea

Notice that $h$ here contains an extra term compared with all of the
previous examples considered. This means that in principle it can have
three, rather than two, zeroes and so there can be as many as three
horizons. This allows the solutions to have a more complex causal
structure than before, allowing in particular examples having a causal
structure similar to that of an RNdS black hole.

\paragraph{Class V$_i$.}
This class of solutions assumes $q = 0$ and leads to non vanishing
$k$ and three terms in the potential for the dilaton. We have the
following conditions
\be
\label{class3v}
\left\{
\begin{array}{l}
\lambda_1 \rho \sqrt{n} = 2\,, \\ [3pt]
(\sigma + \lambda_2)\,\rho  = 2\sqrt{n} \,,  \\ [3pt]
\alpha \lambda_3 \sqrt{n}=  2 \beta \rho  \,, \\[3pt] 
\Lambda_1 = \frac{\alpha\,\lambda_3\,n(n-1)\,k}{\lambda_3-\lambda_1}\,, \\[3pt]
\eta\,Q^2 [\sigma + \lambda_3 ] = \Lambda_2 [\lambda_2 - \lambda_3]\,.
\end{array} \right.
\ee
These imply the constraints $\lambda_1= ({\sigma + \lambda_2})/{n}$
and $\lambda_3 = {4\beta}/{\alpha\lambda_1 n} = {4\beta
  }/{\alpha ( \sigma + \lambda_2)}$.  The function $h$ of the
metric, using these constraints, becomes
\bea
\label{intsolv}
 h(r)&=&-2 M r^{1-n +\beta\rho^2/\alpha} -
\frac{\lambda_1\Lambda_1 \, r^{2\beta\rho^2/\alpha}}
{2\beta\rho \sqrt{n} [n- 1 +\beta\rho^2/\alpha]}- 
\nonumber \\
&& - \frac{\Lambda_3\, r^{2}}{\alpha \,n\,[n +1
-\beta\rho^2/\alpha]} 
+ \frac{[\eta\,\sigma\,Q^2 - \lambda_2 \Lambda_2] \,
r^{2+2\beta\rho^2/\alpha - \lambda_2\rho \sqrt{n}}}
 {2\beta\rho \sqrt{n} [n+ 1 + \beta\rho^2/\alpha -
                      \lambda_2 \rho \sqrt{n}]}\,.
\eea
These geometries also can have at most three horizons with new
interesting properties, as we illustrate below.  There is always a
singularity at the origin (as is also true for the Liouville potential
considered earlier).

Notice that these solutions include de Sitter and anti-de Sitter
space, corresponding to the choice of a constant scalar, $\phi =
\phi_m$, sitting at a stationary point of the potential. This is a new
feature which arises because the potential is now complicated enough
to have maxima and minima.

\subsubsection{Solutions with three horizons}

In this subsubsection we focus on two interesting examples of
solutions with a potential given by the sum of two or more
exponentials. We choose these examples to have \emph{three} horizons
to illustrate their difference from the geometries obtained using the
Liouville potential, which had at most two horizons. In particular, we
find a static solution with three horizons, previously unknown, and
with potentially interesting cosmological applications.

Consider for these purposes point-like objects in 5 dimensions (i.e.\
$n=3$ and $q=0$). Without loss of generality we may take the kinetic
parameters to be canonically normalized: $\alpha=1$, $\beta = \eta =
{1}/{2}$. We leave free the values of the conformal couplings and
the terms in the Liouville terms, and so also of $\rho$. Since we are
interested only in the global properties of the space-times, we give
here only the expression for the metric and not the other fields.

{\renewcommand\belowcaptionskip{1em}
\FIGURE[t]{\epsfig{file=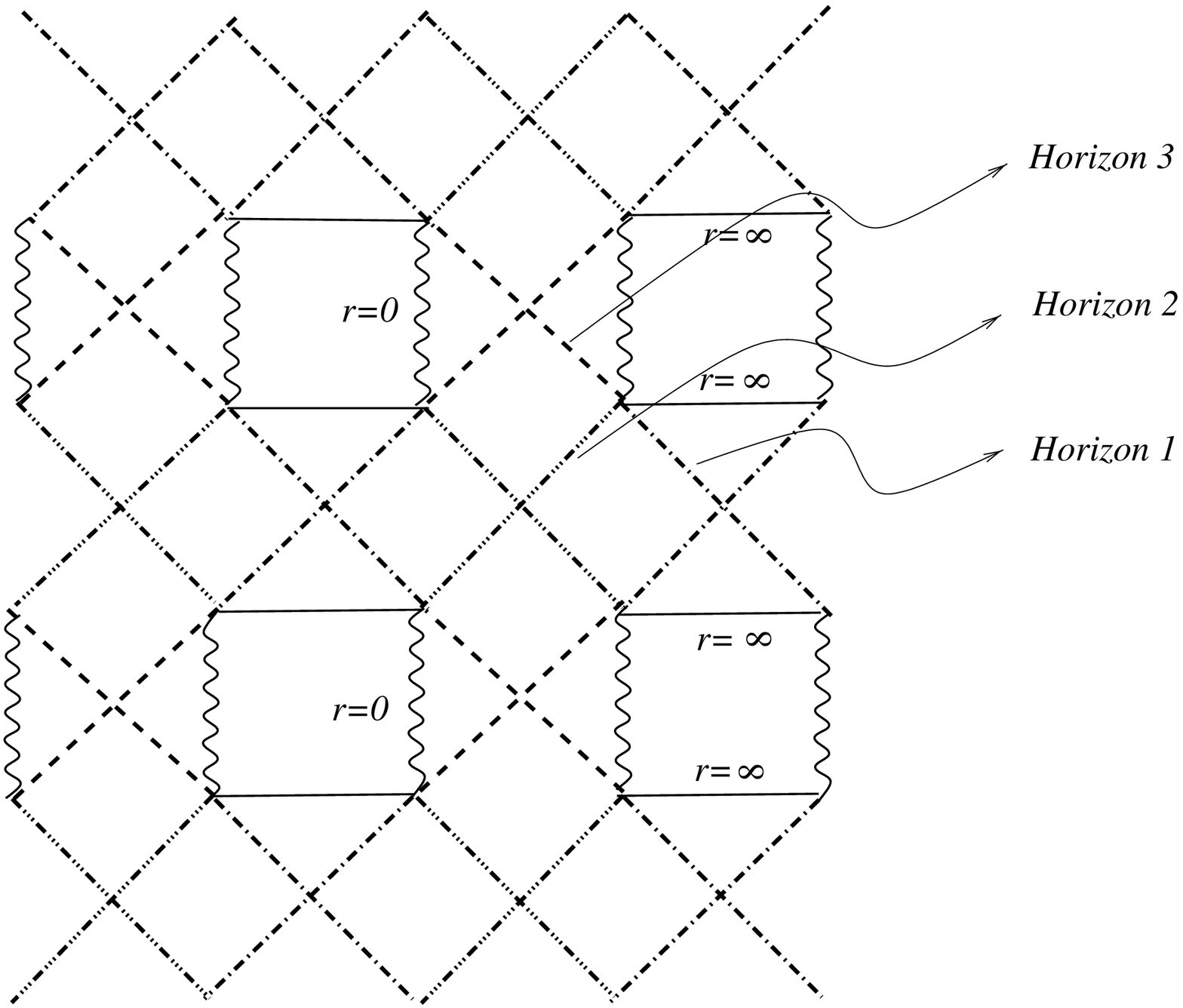, width=.72\textwidth}%
\caption{Penrose diagram for the example~(\ref{3horis1}) having up to
  three horizons with the outer region being
  time-dependent.\label{dSRN}}}
}

\paragraph{Class $IV_i$.}
The metric in this case is given by
\be
ds^2 =  -h(r) dt^2 +\frac{r^{\rho^{2}}}{h(r)}\,dr^2 + r^2 d\Omega_{k,3}^2\,,
\ee
where $h$ is
\bea
\label{3horis1}
h(r) = -2 M r^{\frac{\rho^{2}}{2}-2}
-\frac{2\lambda_{1}\Lambda_{1}\, r^{\rho^{2}} }{\rho\sqrt{3}(4+\rho^{2})}
-\frac{2 \Lambda_{2}r^{2} }{3(8-\rho^{2})}
+\frac{Q^2}{3(4+\rho^{2})r^{4}}\,.
\eea
In this case we can have three horizons only for spacetimes for which
$h(r)$ is negative for large $r$, and so which are asymptotically
time-dependent. These have the causal structure of a de
Sitter-Reissner-Nordstr\"om black hole (see figure~\ref{dSRN}). This
kind of geometry may be arranged by any of the following choices:
\begin{itemize}
\item If $\rho^{2}>8$ the geometry has three horizons whenever
  $\Lambda_{1}$ is positive, while $M$ and $\Lambda_{2}$ are negative.
\item If $2<\rho^{2}<6$ we have three horizons when $\Lambda_{1}$ and
  $M$ are both positive, while $\Lambda_{2}$ is negative.
\item If $\rho^{2}<2$ we require $\Lambda_{2}$ and $M$ both positive
  and $\Lambda_{1}$ negative.
\end{itemize}

\paragraph{Class $V_i$.}
The metric in this case becomes
\be
ds^2 =  -h(r) dt^2 +\frac{r^{\rho^{2}}}{h(r)}\,dr^2 + r^2 d\Omega_{k,3}^2\,,
\ee
where $h$ is given by
\be\label{3horis2}
h(r) = -2 M r^{\frac{\rho^{2}}{2}-2}
-\frac{2\lambda_{1}\Lambda_{1}\,r^{\rho^{2}}}{\rho\sqrt{3}(4+\rho^{2})}
-\frac{2 \Lambda_{3}\,r^{2}}{3(8-\rho^{2})} 
 +\frac{[\sigma\,Q^2-
         2\lambda_{2}\Lambda_{2}]\,r^{\rho^{2}+2-\sqrt{3}\lambda_{2}\rho}}
               {\rho\,\sqrt{3}(8+\rho^{2}-2\sqrt{3} \lambda_{2} \rho)} \,.
\ee
\FIGURE[t]{\epsfig{file=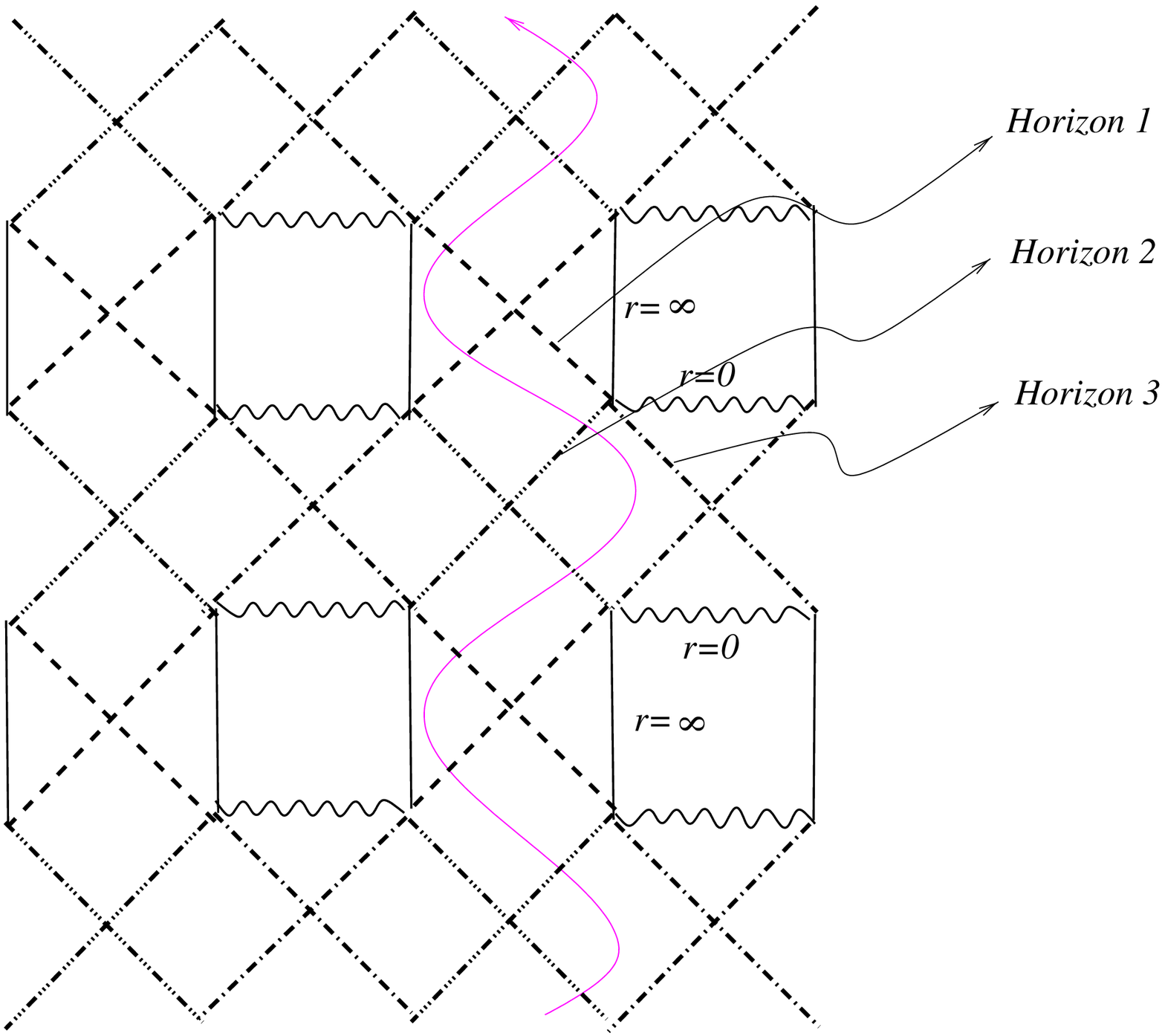, width=.72\textwidth}%
\caption{Penrose diagram for the example in eq.~(\ref{3horis2})
where there can be up to three horizons with the outermost region
being static. Notice that an observer can pass from one Universe
to another \emph{without} crossing a dangerous Cauchy horizon.
\label{SchadSRN}}}
In this case, we can have three horizons both for asymptotically
time-dependent and asymptotically static solutions. They either have
the causal structure of a de Sitter-Reissner-Nordstr\"om black hole
(see figure~\ref{dSRN}), or a new structure represented by
figure~\ref{SchadSRN}. In this case we obtain asymptotically static
solutions with three horizons (see figure~\ref{SchadSRN}) under the
following circumstances.
%\begin{itemize}
%\item If $\lambda_{2}$ and $\rho$ are both positive, then:
%\begin{itemize}
%\item For $\rho^{2} \, > \,8$ and $\lambda_2\rho\sqrt{3}<2$,
%  $\Lambda_{1}$ and $M$ must both be negative, $\Lambda_3$ be
%  positive, and $Q^{2} \, > \, 2\lambda_{2}\Lambda_{2}$.
%\item For $2<\rho^{2} <8$ and $\lambda_2\rho\sqrt{3}<2$ $\Lambda_{3}$
%  and $M$ must be positive, $\Lambda_{1}$ negative, and $Q^{2} \,> \,
%  2\lambda_{2}\Lambda_{2}$.
%\item For $\lambda_2\rho\sqrt{3}<\rho^{2}<2$, $\Lambda_{1}$, $M$ and
%  $\Lambda_{3}$ must be positive and $Q^{2} \,> \,
%  2\lambda_{2}\Lambda_{2}$.
%\item For $\rho^{2}<\lambda_2\rho\sqrt{3}<2 $, $\Lambda_{1}$ and $M$
%  must be positive, $\Lambda_{3}$ negative and $Q^{2} \,< \,
%  2\lambda_{2}\Lambda_{2}$.
%\end{itemize}
%\item If $\lambda_{2}$ is positive and $\rho$ negative, then we need
%  $Q^{2} \,> \, 2\lambda_{2}\Lambda_{2}$ and:
%\begin{itemize}
%\item For $\rho^{2} \, >\,8$, $\Lambda_{1}$ and $M$ must be negative,
%  while $\Lambda_{3}$ is positive.
%\item For $8>\rho^{2} \, >\,2$, $\Lambda_{1}$ and $\Lambda_{3}$ must
%  be negative and $M$ positive.
%\item For $\rho^{2} <2$, $\Lambda_{3}$, $M$ and $\Lambda_{1}$ must all
%  be positive.
%\end{itemize}
%\end{itemize}
\begin{itemize}
\item If $\lambda_{2}$ and $\rho$ are both positive, then:
\begin{itemize}
\item For $\rho^{2}  > 8$ and $\lambda_2\rho\sqrt{3}<2$,
$\Lambda_{1}$ and $M$ must both be negative, $\Lambda_3$ be
positive, and $\sigma Q^{2} -  2\lambda_{2}\Lambda_{2} > 0 $.
\item For $2<\rho^{2} <8$ and $\lambda_2\rho\sqrt{3}<2$,
$\Lambda_{3}<0$,  $M$ and $\Lambda_{1}$ must be positive,
and $\sigma   Q^{2} - 2\lambda_{2}\Lambda_{2} > 0$.
\item For $\lambda_2\rho\sqrt{3}<\rho^{2}<2$, $\Lambda_{1}$, $M$ and
$\Lambda_{3}$ must be positive and $\sigma   Q^{2} -
2\lambda_{2}\Lambda_{2}>0 $.
\item For $\rho^{2}<\lambda_2\rho\sqrt{3}<2 $, $\Lambda_{1}$
and $M$ must be positive, $\Lambda_{3}$ negative and
$\sigma   Q^{2} - 2\lambda_{2}\Lambda_{2} < 0$.
\end{itemize}
\item If $\lambda_{2}$ is positive and $\rho$ negative, then
we need $\sigma Q^{2} - 2\lambda_{2}\Lambda_{2}>0$ and:
\begin{itemize}
\item For $\rho^{2}   > 8$, $\Lambda_{1}$ and $M$ must be negative,
while
$\Lambda_{3}$ is positive.
\item For $8>\rho^{2}    > 2$, $\Lambda_{1}$ and $\Lambda_{3}$
must be negative and $M$ positive.
\item For $\rho^{2} <2$, $\Lambda_{3}$,
$M$ and $\Lambda_{1}$ must all be positive.
\end{itemize}
\end{itemize}

\subsection{A gaussian potential}

We now consider a different example which leads to a gaussian scalar
potential. Proceeding as for the previous examples we start by
choosing a suitable \ansatz\ for the dilaton function $S(\ln r)$. To
illustrate the method we restrict ourselves to the simple case of
$k=Q=0$ (which also implies $c=1$, see eq.~(\ref{constrayygen})). In
this case the functions $F_1$ and $F_2$ become:\footnote{In this
  section we will choose the values $\alpha=\beta =1/2$.}
\be
\label{efe}
 F_1 = -\frac{\xi}{\mathcal M^2} e^{f+L}\, V(S)\,,
\qquad F_2 = \frac{\xi}{2\mathcal M^2} e^{f+L} \frac{dV}{dS} 
\ee
where we now reserve the symbol $'$ to only denote $d/df$. This
implies that
\be 
S' F_2 = -\frac{1}{2} \left[ F_1'-\left(1+L'\right) F_1\right] 
\ee
and so equation~(\ref{sdev}) becomes the following
integro-differential equation for the function $F_1(f)$:
\be 
-S'S''e^{-\mathcal M^2
f}\int e^{\mathcal M^2f} F_1 df \ =\ S'^2F_1+\frac{1}{2}
\left(F_1'- \left(1+L'\right) F_1 \right). 
\ee
Notice that we absorb the integration constant, $2M$, into the
integral without loss of generality. Defining
\be
\label{gama} 
\Gamma \equiv \int e^{\mathcal M^2f} F_1 \, df
\ee
the above equation becomes a differential equation for the
function $\Gamma(f)$:
\be 
\Gamma'' - \left(\mathcal
M^2+1-L'\right) \Gamma' + L'' \Gamma = 0 
\ee
where we use $L'=S'^2$ and therefore $L''=2S' S''$. This is a simple
second-order differential equation for $\Gamma$, which can be solved
given simple functional forms for $L(f)$. Its solution then determines
$\Gamma$ and so also the scalar potential $V$. In particular the case
$L'=$ constant reproduces the exponential potential discussed earlier,
as expected.

A less trivial possibility is obtained by taking
\be 
S = e^{-\gamma f} 
\ee
which implies $L'=\gamma^2 e^{2\gamma f}$, $L''=-2\gamma^3 \,
e^{2\gamma f}$. If the exponent $\gamma$ is fixed to
\be 
\gamma = \frac{1}{2} \left(\mathcal M^2 +1 \right) 
\ee
the differential equation becomes
\be H' = -\gamma^2 e^{-2\gamma f} H 
\ee 
for 
\be 
H \equiv   \Gamma'- \left(\mathcal M^2+1\right) \Gamma 
\ee
which implies
\be 
\Gamma'- \left(\mathcal M^2+1\right) \Gamma = K \,
{\exp}\left({\frac{\gamma}{2} e^{-2\gamma f}}\right) 
\ee
where $K$ is an integration constant. This last equation is also
easily solved to give
\be 
\Gamma = K e^{2\gamma f}\ \int^f e^{-2\gamma \tilde
f}\exp\left({\frac{\gamma}{2}e^{-2\gamma \tilde f}}\right) d\tilde
f = -\gamma^{-2}e^{-2\gamma f} + 2M \, . 
\ee

From this expression we can use the definition of $\Gamma$ to find the
scalar potential as a function of $S= e^{-\gamma f}$ using
equations~(\ref{efe}) and~(\ref{gama}):
\be 
V(S) = \tilde K  e^{\gamma S^2}\left[ \left(2-\gamma S^2\right) -4M
e^{-\frac{\gamma}{2} S^2}\right]
\ee
Where $\tilde K$ is another integration constant. Finally, knowing the
potential and the functional dependence of $S(f)$ we can determine the
metric function $h(r)$ using~(\ref{ache}) and obtain the geometry
corresponding to this potential:
\be 
h(r) = -f' r e^L\ =  e^{-\mathcal M^2 f} \Gamma r
e^L=r^{1-\mathcal M^2}{\rm e}^{-\frac{\gamma}{2}r^{-2\gamma}}\left(2M
-\gamma^{-2}{\exp}\left({\frac{\gamma}{2}r^{-2\gamma}}\right)\right) \, .
\ee
Here we reintroduce the integration constant $2M$.

Notice that all of the unknown functions are now determined (recall
$r=e^f$). From here we may find the causal properties of the geometry
by finding the singularity and zeroes of the function $h(r)$. In this
case if $M > 0$ we have only one horizon. Since $h(r)\rightarrow
-\infty$ when $r\rightarrow 0$ we see that the singularity at $r = 0$
is timelike. Furthermore for $r \rightarrow \infty$ we have
$h(r)\rightarrow 0$. The Penrose diagram for this geometry has then
the form of the S-brane of figure~\ref{sbrpen}.

Clearly this section just scratches the surface of the method by
providing a few simple illustrations. Even though the technique is
practical only for solving the inverse problem --- \emph{ie} finding
the dilaton potential which generates a given \ansatz\ for the
geometry --- it is remarkable to be able to obtain explicit solutions
for such complicated potentials. More complicated options are
possible, such as the choice $L'=f^m$ which leads to Sturm-Liouville
systems whose solutions can be represented by special functions. (In
particular for $m=1$ the potential is a combination of exponentials
and the error function.)

\section{Conclusions}\label{part5}

With this paper we begin the search for solutions to the
Einstein-scalar-antisymmetric-form system in the presence of a
non-vanishing scalar potential, and describe solutions which are both
static and time-dependent for asymptotically large coordinates. The
solutions we find have $(q+1)$-dimensional singularities which act as
sources for the fields, and our solutions generalise the black
$q$-brane and S-brane solutions which were previously known for the
case of vanishing scalar potential.

\pagebreak[3] 

We obtain our solutions by solving the field equations subject to an
\ansatz\ for the metric and dilaton fields. By treating our \ansatz\
for a general scalar potential we reduce the problem of generating
solutions to the problem of integrating a single, non-linear
differential equation for the scalar field, which in principle can be
done numerically for specific potentials. The solution of this
equation provides the global structure of the spacetime, including the
singularities, horizons and asymptotic behavior.

\looseness=1We apply this approach to simple examples, including the practical
case of the Liouville potential which arises in many gauged
supergravity models. We also consider potentials which are sum of
exponentials, and to illustrate of the power of the technique we
briefly consider some more complicated potentials. The spacetimes
obtained in this way have both static and time-dependent metrics in
their asymptotic regions, and as such provide interesting starting
points for building novel cosmological scenarios. Among the possible
applications are the description of bouncing (or cyclic) universes
without unstable matter content or quintessence-like accelerating
universes which are not asymptotically flat.

Our solutions contain as some particular cases, examples which were
considered previously in the literature~\cite{mann,cai,reall}, and we
verify that our solutions reduce to these in the appropriate corners
of parameter space.

\looseness=1Solutions using the Liouville potential have immediate application to
known massive and gauged supergravities in various dimensions, and we
find new solutions for these systems by specializing to most of the
known gauged supergravities in different numbers of
dimensions. Besides generating solutions we find relationships among
many of them by uplifting them to 10 dimensions and applying various
duality transformations. Some of these geometries preserve half the
model's supersymmetries in an extremal limit consisting of vanishing
mass and charge. These new configurations may play an interesting role
in the further understanding the vacua of these theories.  Notice also
that other potentially interesting solutions can be obtained from
those we present here, either by analytic continuation or by use of
the various duality symmetries, in the same spirit as
in~\cite{bqrtz,bmq}.

We do not know how to embed some of our solutions into
higher-dimensional supergravities and for these solutions we do not
have a proper string pedigree. Among the solutions we find in this
class are solutions to 6-dimensional Salam-Sezgin supergravity, and
solutions using scalar potentials which are the sum of exponentials. A
better understanding of the physical relevance of these configurations
is an interesting open question which we leave for the future.

\acknowledgments

We would like to thank R.~Gibbens, G.~Gibbons, M.~Schvellinger and
T.~Tran for valuable discussions.  C.B. is indebted to D.A.M.T.P. at
Cambridge for their hospitality during an early part of this work, and
to McGill University, N.S.E.R.C. (Canada) and F.C.A.R. (Qu\'ebec) for
research funding. C. N. is supported by a Pappalardo Fellowship, and
in part by funds provided by the U.S. Department of Energy (D.O.E.)
under cooperative research agreement \# DF-FC02-94ER40818.
F.~Q.~thanks CERN for hospitality and PPARC for partial financial
support. G.~T.~is supported by the European TMR Networks
HPRN-CT-2000-00131, HPRN-CT-2000-00148 and
HPRN-CT-2000-00152. I.~Z.~C.~ is supported by CONACyT, Mexico.

\pagebreak[3] 

\appendix

\section{Analysis of all cases for single Liouville potential}
\label{appendix}

In this appendix we report on the geometrical structure of the three
classes of solutions studied in the paper. We consider only the cases
with non-trivial dilaton here, but the constant dilaton case can be
obtained from the equations of motion straightforwardly. For
simplicity and clarity, we concentrate the present appendix on in the
case $q=0$ (and so $\M^2 =n$).

\paragraph{Class I.}
This class of solutions are defined for zero spatial curvature $k=0$.
We recall here the form of the metric, which is given
by~(\ref{class1}). The metric in this case becomes
\be
h(r) = -2M \frac{r^{\beta\rho^2/\alpha}}{r^{\M^2 -1}} - \frac{\Lambda\,
 r^2}{\alpha\,\M^2[\M^2-\beta\rho^2/\alpha +1]}+
+ \frac{\eta\,Q^2}{\alpha\,\M^2[2n-\M^2-1+ \beta\rho^2/\alpha]
\,r^{2(n-1)}}\,,
\ee
and $g(r)=h(r)\,r^{-2\beta\rho^2/\alpha}$.  The dilaton and gauge
fields are given by~(\ref{dilaton}),~(\ref{qform}) with the relevant
values of the parameters.  One common feature of all the solutions is
that all of them have a curvature singularity at $r=0$. Apart from
this, we have the following cases:

\begin{description}
\item[a)]$ \beta\rho^2 > \alpha\,(\M^2 +1)$.

${\bf M>0}$  and $\Lambda< 0$ or $\Lambda>0$ .

First of all notice that the sign in the $\Lambda$ term changes.  In
both cases, we can have at most one Cauchy horizon and the most outer
region is time-dependent\footnote{We will always call the kind of
  solutions, where the most outer region is time dependent,
  \emph{cosmological} and those where it is static,
  \emph{static}.}. It is interesting to notice that in this case, the
asymptotic infinity is still null-like~\cite{wald}. Then the Penrose
diagram looks like that of the S-brane (see figure~\ref{sbrpen}).

${\bf M<0}$

For $\Lambda>0$, the solution is static everywhere and there are no
horizons at all. There is a naked singularity at the origin and the
asymptotic infinity is null-like.

For $\Lambda<0$ instead, the solution is static and there can be up to
two regular horizons. The geometry is asymptotically flat with a
Reissner-Nordstr\"om-like black hole Penrose diagram.

\item[b)]$\alpha<\beta\rho^2 \le \alpha \,(\M^2+1)$.

${\bf M>0}$.

For $\Lambda>0$, we again have a cosmological solution with the same
geometry as the S-brane, being asymptotically flat at infinity.

For $\Lambda<0$, we have a richer structure. We will have a static
solution (defined by the most outer region) and at most two
horizons. The causal structure of the solution will look exactly as
that of the Reissner-Nordstr\"om black hole.

${\bf M<0}$.

For $\Lambda>0$ the geometry is like in the positive mass case.

For $\Lambda<0$ the geometry is everywhere static with a naked
singularity at the origin, as in the previous positive mass case.

\item[c)]$\beta\rho^2 < \alpha$.

${\bf M>0}$.

For $\Lambda>0$, we have a cosmological solution with a Cauchy
horizon, but now the asymptotic infinity is not null-like but
space-like. The Penrose diagram is like a dS-S-brane solution (see
figure~\ref{noflat}).

For $\Lambda<0$, the solutions are static and can have two horizons
and, being not asymptotically flat, but AdS-like. The Penrose diagram
looks like an AdS-Reissner-Nordstr\"om black hole.

${\bf M<0}$.

For $\Lambda>0$ and $\Lambda<0$, this case reduces to the positive
mass case above.
\end{description}

\paragraph{Class II.}
These solutions are defined for non zero spatial curvature $k=-1,1$.
The for of the metric is given by (for $q=0$)~(\ref{class2}),
\be
h(r) = -2 M \frac{r^{\beta\rho^2/\alpha}}{r^{\M^2 -1}} -
\frac{\lambda\,\Lambda\,r^{2\beta\rho^2/\alpha}}{2\beta\rho\,\M
[\M^2-1+\beta\rho^2/\alpha]}
+\frac{\eta\,Q^2}{\alpha\,\M^2 [2n -\M^2+\beta\rho^2/\alpha-1]\,r^{2(n-1)}}\,,
\ee
and $g(r)=h(r)\,r^{-2\beta\rho^2/\alpha}$ again.  The dilaton and
gauge fields are given by~(\ref{dilaton}),~(\ref{qform}) with the
relevant values of the parameters.  Again, all the solutions have a
curvature singularity at $r=0$. Notice that the sign of the
cosmological constant depends on the sign if the spatial curvature
(see class II solutions in the text). We can have the following cases:

\begin{description}
\item[a)]$\beta\rho^2 \ge \alpha$.

${\bf M>0}$.

For $\Lambda>0$, or $k=1$, the solutions are time dependent and there
can only be a regular horizon and the singularity at the origin will
be time-like. The solutions are not asymptotically flat, nor dS but
the asymptotic infinity is space-like, like in a dS case. Then, the
Penrose diagram looks like the dSS-brane (see figure~\ref{noflat}).

For $\Lambda<0$, or $k=-1$, there can be at most two regular horizons
and the solution is static. The singularity in the origin is time-like
and the asymptotic infinity is time-like. Then the Penrose diagram is
like a RNadS black hole (see figure~\ref{adsrn}).

${\bf M<0}$.

For $\Lambda>0$, and $k=1$, the solutions are time dependent
everywhere with a space-like initial singularity at the origin.

For $\Lambda<0$, and $k=-1$ the solutions are static everywhere with a
naked time-like singularity at the origin.

\item[b)]$\beta\rho^2 < \alpha$.

${\bf M>0}$.

For $\Lambda>0$, $k=-1$, we have again a situation very similar to the
previous case, (a), but now the Penrose diagram is like the S-brane
one.

For $\Lambda<0$, $k=1$, the structure is very similar to case (a),
except that the solutions now have a null-like asymptotic infinity.
So the Penrose diagram looks like the pure Reissner-Nordstr\"om black
hole.

${\bf M<0}$.

For $\Lambda>0$, and $k=-1$ we have a situation like the positive mass
case, where the Penrose diagram is that of the S-brane.

For $\Lambda<0$, and $k=1$, the solutions are static everywhere like
in the previous case.
\end{description}

\paragraph{Class III.}
These solutions are defined only for positive spatial curvature $k=1$.
The metric is given by~(\ref{intsol}),
\bea
h(r)= -2 M \frac{r^{\beta\rho^2/\alpha}}{r^{\M^2 -1}} -
\frac{\Lambda\,r^2}{\alpha\,\M^2[\M^2+1-\beta\rho^2/\alpha]}
+ \frac{\sigma\eta \,Q^2\,r^{2\beta\rho^2/\alpha}}
{2\beta\rho\M[\M^2-1+\beta\rho^2/\alpha]}\,,
\eea
 $g(r)=h(r)\,r^{-2\beta\rho^2/\alpha}$.
 All the solutions  have a
curvature singularity at $r=0$.  So we can have the following cases:

\begin{description}
\item[a)]$\beta\rho^2 > \alpha(\M^2 +1)$.

${\bf M>0}$.

For $\Lambda>0$, the solution is static and there may be up to two
horizons. The singularity is time-like and the geometry is like that
of a RNadS black hole (see figure~\ref{adsrn}).

For $\Lambda>0$, we can have at most one regular horizon and the
solution is static. There is a space-like singularity at the origin
and the asymptotic infinity is time-like. Then the Penrose diagram
looks like that of an adS-Schwarszchild black hole (see
figure~\ref{noflat}).

${\bf M<0}$.

For $\Lambda>0$, the solution is static everywhere with a naked
singularity at the origin.

For $\Lambda<0$, the solution is static with at most one horizon. The
Penrose diagram is like that of a adS-Schwarszchild black hole (see
figure~\ref{noflat}).

\item[b)]$ \alpha < \beta\rho^2 < \alpha (\M^2 +1)$.

${\bf M>0}$.

For $\Lambda>0$, or $\Lambda<0$, we can have at most one regular
horizon and the solution is static. There is a space-like singularity
at the origin and the asymptotic infinity is time-like. Then the
Penrose diagram looks like that of an AdS-Schwarszchild black hole
(see figure~\ref{noflat}).

\pagebreak[3] 

${\bf M<0}$.

For $\Lambda>0$, there may be at most two regular horizons and a
time-like singularity at the origin. The Penrose diagram looks like a
Reissner-Nordstr\"om-AdS black hole (see figure~\ref{adsrn}).

For $\Lambda<0$, the solution is static without horizons, so there is
a naked time-like singularity at the origin and the asymptotic
infinity will be time-like.

\item[b)]$\beta\rho^2 < \alpha$.

${\bf M>0}$.

This is a very interesting solution for $\Lambda>0$. The solution is
cosmological and there may be up to two regular horizons, one
cosmological and one event horizon. The infinity asymptotic is not
that of de-Sitter space, however it is space-like. The Penrose diagram
looks then, like that of a dS-Schwarschild black hole.  It is also
interesting to note that this is the only case where this structure
comes out, and moreover, there are no cases where the asymptotic
infinity is null-like.

For $\Lambda<0$, is exactly like in the previous case, (a).

${\bf M<0}$.

For $\Lambda>0$, there may be at most one regular horizon and a
time-like singularity at the origin. The Penrose diagram looks like a
dS-S-brane.

For $\Lambda<0$, it coincides with the case (a), for the same value of
$M$.

\item[c)]$\beta\rho^2 = \alpha$.

${\bf M>0}$.

For $\Lambda>0$, there are two possibilities. Either there are no
horizons and the solution is everywhere cosmological with a space-like
asymptotic infinity. Or it may have an event horizon and then it looks
like a AdS-Schwarzschild black hole.

For $\Lambda<0$, the solutions is again like in the first case, (a).

${\bf M<0}$.

For $\Lambda>0$, there are two possibilities. The solutions may be
static everywhere with a naked time-like singularity or there may be
one regular horizon and the structure is then like a dS-S-brane.

For $\Lambda<0$, the solution is the same as in the first case, (a).
\end{description}


\begin{thebibliography}{99} 

\bibitem{pol}
J.~Polchinski, \emph{Dirichlet-branes and Ramond-Ramond charges},
\prl{75}{1995}{4724} [\hepth{9510017}].

\bibitem{gutperle}
M.~Gutperle and A.~Strominger, \emph{Spacelike branes},
\jhep{04}{2002}{018} [\hepth{0202210}].

\bibitem{lukas}
A.~Lukas, B.A. Ovrut and D.~Waldram, \emph{Cosmological solutions of
  type-II string theory}, \plb{393}{1997}{65} [\hepth{9608195}];\\
H.~Lu, S.~Mukherji, C.N. Pope and K.W. Xu, \emph{Cosmological
  solutions in string theories}, \prd{55}{1997}{7926}
[\hepth{9610107}];\\
A.~Lukas, B.A. Ovrut and D.~Waldram, \emph{String and M-theory
  cosmological solutions with ramond forms}, \npb{495}{1997}{365}
[\hepth{9610238}];\\
H.~Lu, S.~Mukherji and C.N. Pope, \emph{From p-branes to cosmology},
\ijmpa{14}{1999}{4121} [\hepth{9612224}].

\bibitem{sbranes1}
C.-M. Chen, D.V. Gal'tsov and M.~Gutperle, \emph{S-brane solutions in
  supergravity theories}, \prd{66}{2002}{024043} [\hepth{0204071}].

\bibitem{sbranes2}
M.~Kruczenski, R.C. Myers and A.W. Peet, \emph{Supergravity S-branes},
\jhep{05}{2002}{039} [\hepth{0204144}];\\
S.~Roy, \emph{On supergravity solutions of space-like D$p$-branes},
\jhep{08}{2002}{025} [\hepth{0205198}];\\
A.~Buchel, P.~Langfelder and J.~Walcher, \emph{Does the tachyon
  matter?}, \ap{302}{2002}{78} [\hepth{0207235}].

\bibitem{townsend2003}
P.K. Townsend and M.N.R. Wohlfarth, \emph{Accelerating cosmologies
  from compactification}, \prl{91}{2003}{061302} [\hepth{0303097}].

\bibitem{accelerating}
N.~Ohta, \emph{Accelerating cosmologies from S-branes},
\prl{91}{2003}{061303} [\hepth{0303238}];\\
S.~Roy, \emph{Accelerating cosmologies from m/string theory
  compactifications}, \plb{567}{2003}{322} [\hepth{0304084}];\\
M.N.R. Wohlfarth, \emph{Accelerating cosmologies and a phase
  transition in M-theory}, \plb{563}{2003}{1} [\hepth{0304089}];\\
R.~Emparan and J.~Garriga, \emph{A note on accelerating cosmologies
  from compactifications and s-branes}, \jhep{05}{2003}{028}
[\hepth{0304124}];\\
N.~Ohta, \emph{A study of accelerating cosmologies from superstring/m
  theories}, \ptp{110}{2003}{269} [\hepth{0304172}];\\
C.-M. Chen, P.-M. Ho, I.P. Neupane and J.E. Wang, \emph{A note on
  acceleration from product space compactification},
\jhep{07}{2003}{017} [\hepth{0304177}];\\
M.~Gutperle, R.~Kallosh and A.~Linde, \emph{M/string theory, s-branes
  and accelerating universe}, \emph{JCAP} {\bf 0307} (2003) 001
[\hepth{0304225}].

\bibitem{ohta}
N.~Ohta, \emph{Intersection rules for s-branes}, \plb{558}{2003}{213}
[\hepth{0301095}].

\bibitem{gqtz}
C.~Grojean, F.~Quevedo, G.~Tasinato and I.~Zavala, \emph{Branes on
  charged dilatonic backgrounds: self-tuning, Lorentz violations and
  cosmology}, \jhep{08}{2001}{005} [\hepth{0106120}].

\bibitem{kounnas}
L.~Cornalba and M.S. Costa, \emph{A new cosmological scenario in
  string theory}, \prd{66}{2002}{066001} [\hepth{0203031}];\\
L.~Cornalba, M.S. Costa and C.~Kounnas, \emph{A resolution of the
  cosmological singularity with orientifolds}, \npb{637}{2002}{378}
[\hepth{0204261}].

\bibitem{bqrtz}
C.P. Burgess, F.~Quevedo, S.J. Rey, G.~Tasinato and I.~Zavala,
\emph{Cosmological spacetimes from negative tension brane
  backgrounds}, \jhep{10}{2002}{028} [\hepth{0207104}].

\bibitem{talk}
F.~Quevedo, G.~Tasinato and I.~Zavala, \emph{S-branes, negative
  tension branes and cosmology}, \hepth{0211031}.

\bibitem{bmqtz}
C.P. Burgess, P.~Martineau, F.~Quevedo, G.~Tasinato and I.~Zavala~C.,
\emph{Instabilities and particle production in s-brane geometries},
\jhep{03}{2003}{050} [\hepth{0301122}].

\bibitem{cc}
L.~Cornalba and M.S. Costa, \emph{On the classical stability of
  orientifold cosmologies}, \cqg{20}{2003}{3647} [\hepth{0302137}].

\bibitem{wilt1}
D.L. Wiltshire, \emph{Global properties of Kaluza-Klein cosmologies},
\prd{36}{1987}{1634}.

\bibitem{mann}
K.C.K. Chan, J.H. Horne and R.B. Mann, \emph{Charged dilaton black
  holes with unusual asymptotics}, \npb{447}{1995}{441}
[\grqc{9502042}].

\bibitem{cai}
R.-G. Cai, J.-Y. Ji and K.-S. Soh, \emph{Topological dilaton black
  holes}, \prd{57}{1998}{6547} [\grqc{9708063}].

\bibitem{reall}
H.A. Chamblin and H.S. Reall, \emph{Dynamic dilatonic domain walls},
\npb{562}{1999}{133} [\hepth{9903225}].

\bibitem{charmousis}
C.~Charmousis, \emph{Dilaton spacetimes with a Liouville potential},
\cqg{19}{2002}{83} [\hepth{0107126}].

\bibitem{langlois}
D.~Langlois and M.~Rodriguez-Martinez, \emph{Brane cosmology with a
  bulk scalar field}, \prd{64}{2001}{123507} [\hepth{0106245}].

\bibitem{zhang}
R.-G. Cai and Y.-Z. Zhang, \emph{Black plane solutions in
  four-dimensional spacetimes}, \prd{54}{1996}{4891}
[\grqc{9609065}];\\
R.-G. Cai, Y.S. Myung and Y.-Z. Zhang, \emph{Check of the mass bound
  conjecture in de~Sitter space}, \prd{65}{2002}{084019}
[\hepth{0110234}].

\bibitem{paul}
P.K. Townsend, \emph{Quintessence from M-theory}, \jhep{11}{2001}{042}
[\hepth{0110072}].

\bibitem{boonstra}
H.J. Boonstra, K.~Skenderis and P.K. Townsend, \emph{The domain
  wall/QFT correspondence}, \jhep{01}{1999}{003} [\hepth{9807137}].

\bibitem{behrndt}
K.~Behrndt, M.~Cveti\v{c} and W.A. Sabra, \emph{Non-extreme black
  holes of five dimensional $N=2$ AdS supergravity},
\npb{553}{1999}{317} [\hepth{9810227}].

\bibitem{cvetic99}
M.~Cveti\v{c} et al., \emph{Embedding AdS black holes in ten and
  eleven dimensions}, \npb{558}{1999}{96} [\hepth{9903214}].

\bibitem{bowcock}
P.~Bowcock, C.~Charmousis and R.~Gregory, \emph{General brane
  cosmologies and their global spacetime structure},
\cqg{17}{2000}{4745} [\hepth{0007177}].

\bibitem{wilt2}
S.J. Poletti, J.~Twamley and D.L. Wiltshire, \emph{Charged dilaton
  black holes with a cosmological constant}, \prd{51}{1995}{5720}
[\hepth{9412076}];\\
D.L. Wiltshire, \emph{Dilaton black holes with a cosmological term},
\emph{J.  Austral. Math. Soc.} {\bf B41} (1999) 198--216
     [\grqc{9502038}].

\bibitem{cina}
B.~Zhou and C.-J. Zhu, \emph{The complete black brane solutions in
  D-dimensional coupled gravity system}, \hepth{9905146}.

\bibitem{oz}
P.~Brax, G.~Mandal and Y.~Oz, \emph{Supergravity description of
  non-BPS branes}, \prd{63}{2001}{064008} [\hepth{0005242}].

\bibitem{romans}
L.J. Romans, \emph{Massive $N=2a$ supergravity in ten-dimensions},
\plb{169}{1986}{374}.

\bibitem{perry}
A.~Chamblin, M.J. Perry and H.S. Reall, \emph{Non-BPS d8-branes and
  dynamic domain walls in massive IIA supergravities},
\jhep{09}{1999}{014} [\hepth{9908047}].

\bibitem{berg}
E.~Bergshoeff, M.~de~Roo, M.B. Green, G.~Papadopoulos and
P.K. Townsend, \emph{Duality of type-II 7-branes and 8-branes},
\npb{470}{1996}{113} [\hepth{9601150}].

\bibitem{Salam:1984ft}
A.~Salam and E.~Sezgin, \emph{$D = 8$ supergravity},
\npb{258}{1985}{284}.

\bibitem{townew}
P.K. Townsend and P.~van Nieuwenhuizen, \emph{Gauged seven-dimensional
  supergravity}, \plb{125}{1983}{41}.

\bibitem{chamse}
A.H. Chamseddine and W.A. Sabra, \emph{$D = 7 {\rm SU}(2)$ gauged
  supergravity from $D = 10$ supergravity}, \plb{476}{2000}{415}
[\hepth{9911180}].

\bibitem{cpl1}

M.~Cveti\v{c}, H.~Lu and C.N. Pope, \emph{Gauged six-dimensional
  supergravity from massive type-IIA}, \prl{83}{1999}{5226}
[\hepth{9906221}].

\bibitem{romansix}
L.J. Romans, \emph{The F(4) gauged supergravity in six-dimensions},
\npb{269}{1986}{691}.

\bibitem{nunez}
C.~N\'u\~nez, I.Y. Park, M.~Schvellinger and T.A. Tran,
\emph{Supergravity duals of gauge theories from F(4) gauged
  supergravity in six dimensions}, \jhep{04}{2001}{025}
     [\hepth{0103080}].

\bibitem{ss}
H.~Nishino and E.~Sezgin, \emph{Matter and gauge couplings of $N=2$
supergravity in six-dimensions}, \plb{144}{1984}{187};\\
A.~Salam and E.~Sezgin, \emph{Chiral compactification on Minkowski
  $\times S^2$ of $N=2$ Einstein-Maxwell supergravity in
  six-dimensions}, \plb{147}{1984}{47}.

\bibitem{branesphere}
Y.~Aghababaie, C.P. Burgess, S.L. Parameswaran and F.~Quevedo,
\emph{Towards a naturally small cosmological constant from branes in
  6D supergravity}, \hepth{0304256}.

\bibitem{6Dcc}
Y.~Aghababaie, C.P. Burgess, S.L. Parameswaran and F.~Quevedo,
\emph{SUSY breaking and moduli stabilization from fluxes in gauged 6d
  supergravity}, \jhep{03}{2003}{032} [\hepth{0212091}].

\bibitem{romansfive}
L.J. Romans, \emph{Gauged $N=4$ supergravities in five-dimensions and
  their magnetovac backgrounds}, \npb{267}{1986}{433}.

\bibitem{tran}
M.~Schvellinger and T.A. Tran, \emph{Supergravity duals of gauge field
  theories from ${\rm SU}(2)\times \U(1)$ gauged supergravity in five
  dimensions}, \jhep{06}{2001}{025} [\hepth{0105019}].

\bibitem{hull}
C.M. Hull, \emph{A new gauging of $N=8$ supergravity},
\prd{30}{1984}{760};
\emph{Noncompact Gaugings Of N=8 Supergravity,} \plb{142}{1984}{39};
\emph{The minimal couplings and scalar potentials of the gauged $N=8$
  supergravities}, \cqg{2}{1985}{343};\\
C.M. Hull and N.P. Warner, \emph{The structure of the gauged $N=8$
  supergravity theories}, \npb{253}{1985}{650};
\emph{The potentials of the gauged $N=8$ supergravity theories},
\npb{253}{1985}{675}.

\bibitem{linde}
R.~Kallosh, A.D. Linde, S.~Prokushkin and M.~Shmakova, \emph{Gauged
  supergravities, de~Sitter space and cosmology},
\prd{65}{2002}{105016} [\hepth{0110089}].

\bibitem{ahn}
C.-h. Ahn and K.-s. Woo, \emph{Domain wall and membrane flow from
  other gauged $D=4$, $N=8$ supergravity, I}, \npb{634}{2002}{141}
[\hepth{0109010}].

\bibitem{bmq}
C.P. Burgess, R.C. Myers and F.~Quevedo, \emph{On spherically
  symmetric string solutions in four- dimensions}, \npb{442}{1995}{75}
[\hepth{9410142}].

\bibitem{wald}
R.M.~Wald, \emph{General relativity}, The University of Chicago Press,
Chicago 1984.

\end{thebibliography}
\end{document}